\documentclass[reprint, prd, superscriptaddress]{revtex4-1}

\usepackage{graphicx}
\usepackage{amsmath}
\usepackage{amssymb}
\usepackage{amstext}
\usepackage{amsfonts}
\usepackage{braket}
\usepackage{tabularx}
\usepackage{color}
\usepackage[utf8]{inputenc}
\usepackage{hyperref}
\usepackage{natbib}
\usepackage{placeins}

\newcolumntype{Y}{>{\centering\arraybackslash}X}

\begin{document}
\title{Timing the Neutrino Signal of a Galactic Supernova}
\author{Rasmus S. L. Hansen}
\email{rasmus.lundkvist@nbi.ku.dk}
\affiliation{Max-Planck-Institut für Kernphysik, Saupfercheckweg 1
69117 Heidelberg, Germany}
\affiliation{Department of Physics and Astronomy, Aarhus University, Ny Munkegade 120, 8000 Aarhus C, Denmark}
\author{Manfred Lindner}
\email{lindner@mpi-hd.mpg.de}
\affiliation{Max-Planck-Institut für Kernphysik, Saupfercheckweg 1
69117 Heidelberg, Germany}
\author{Oliver Scholer}
\email{scholer@mpi-hd.mpg.de}
\affiliation{Max-Planck-Institut für Kernphysik, Saupfercheckweg 1
69117 Heidelberg, Germany}
\date{\today}

\begin{abstract}
We study several methods for timing the neutrino signal of a Galactic supernova (SN) for different detectors via Monte Carlo simulations. We find that, for the methods we studied, at a distance of $10\,$kpc both Hyper-Kamiokande and IceCube can reach precisions of $\sim1\,$ms for the neutrino burst, while a potential IceCube Gen2 upgrade will reach submillisecond precision. In the case of a failed SN, we find that detectors such as SK and JUNO can reach precisions of $\sim0.1\,$ms while HK could potentially reach a resolution of $\sim 0.01\,$ms so that the impact of the black hole formation process itself becomes relevant. Two possible applications for this are the triangulation of a (failed) SN as well as the possibility to constrain neutrino masses via a time-of-flight measurement using a potential gravitational wave signal as reference.
\end{abstract}

\maketitle
\section{Introduction}
Massive stars above $\sim8M_\odot$ most often end their lives in a great explosion outshining an entire galaxy for a short period of time.
For such core-collapse supernovae (CCSNe), it is predicted that $\sim99\%$ of the released gravitational binding energy is emitted via neutrinos~\citep{jankaneutrinodrivenexplosions, Kotake:2012nd, Mirizzi:2015eza, Horiuchi:2017sku}. State of the art multidimensional simulations model the supernova explosion mechanism as well as the neutrino emission properties~\citep{Glas:2018oyz, Burrows:2019rtd, Muller:2018utr, Vartanyan:2019ssu, Vartanyan:2018iah, Summa:2017wxq, Melson:2015spa}.\\
In the case of a Galactic CCSN, current and near future neutrino detectors will be able to detect thousands of events in a time period of $\sim10$ seconds. The detection of such SN neutrinos offers interesting possibilities.\\
In contrast to photons, neutrinos travel freely through the outer shells of the SN. Therefore in the case of a Galactic CCSN, the neutrino signal will reach us long before any optical signal can be detected.
This way it can serve as an early warning system (see SNEWS~\citep{Antonioli:2004zb}). Besides other methods such as studying the statistics of neutrino-electron elastic scattering~\citep{Tomas:2003xn, Beacom:1998fj}, the precise timing in multiple neutrino detectors can also be used to locate the SN via triangulation~\citep{Beacom:1998fj, Muhlbeier:2013gwa, Brdar:2018zds, Linzer:2019swe,Coleiro:2020vyj}. This is not only important to enable early astronomical observations of the SN, but also to locate it in the case of a failed SN which would not result in any optical signal. In the latter case, locating the SN will allow us to search for and study the SN remnant, potentially observe the progenitors collapse to a black hole as well as to include and study the impact of Earth-matter effects that can be included only if the direction of the neutrino signal is known.
Furthermore, combining neutrino and gravitational wave signals might allow us to determine the mass of neutrinos. This is another application that needs a very precise timing of the neutrino signal.\\
In this work, we present several methods on how characteristic structures of the neutrino signal can be used for precise timing. This is based on simulations of neutrino signals for different detectors using a set of both successful and failed supernova neutrino simulations from the Garching group~\citep{huedepohl_thesis, risetimemasshierarchy}.\\
This paper is structured as follows: In Sec.~\ref{sec2}, we give a short overview on the general neutrino emission properties. In Sec.~\ref{sec3}, we study the neutrino signal in several detectors. In Sec.~\ref{sec4}, we use a Monte Carlo simulation based on the results of Sec.~\ref{sec3} to study several methods for timing the neutrino signal using characteristic structures. Finally, we study two possible applications namely triangulation and the neutrino mass determination in Sec.~\ref{sec5}. Throughout the paper we use natural units $c=\hbar=k_B=1$.

\section{The Neutrino Signal}\label{sec2}
\subsection{Phases of Emission}
\begin{figure}[tbp]
\begin{center}
\includegraphics[width=\columnwidth]{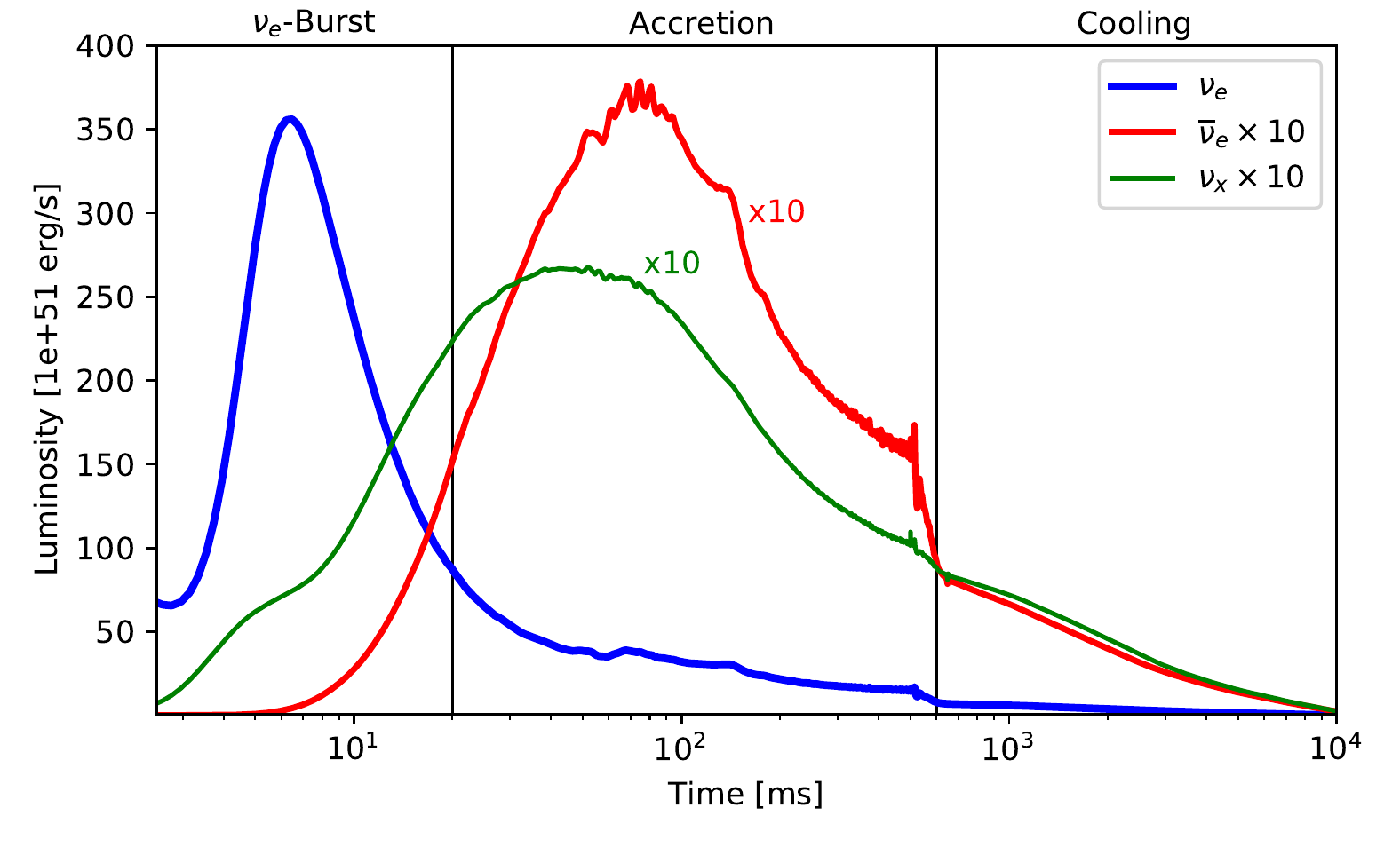}
\caption{{Phases of emission in a typical CCSN taken from the s11.2 cooling model of Hüdepohl~\citep{huedepohl_thesis}
. Note that $\overline{\nu}_e$ and $\nu_x$ are scaled up. One can clearly see the three different phases of emission namely the large $\nu_e$-burst during the first $\sim10\,\text{ms}$, the following accretion phase, and the cooling of the neutron star at the end up to $\sim10\,\text{s}$.}}\label{emissionphases}
\end{center}
\end{figure}

Typically, the neutrino emission from a supernova can be separated into three different phases which are shown in FIG.~\ref{emissionphases}
\begin{enumerate}
\item
$\nu_e$-Burst: During collapse, vast amounts of $\nu_e$ are produced via electron capture. When the collapsing core exceeds a certain density, the neutrinos get trapped inside. Shortly after core bounce, they are suddenly released resulting in a characteristic sharp $\nu_e$-burst with a typical luminosity of $\sim 3.5\cdot10^{53}\,\frac{\text{erg}}{\text{s}}$.
\item
Accretion: Ongoing accretion onto the newly formed protoneutron star produces all types of neutrino flavors via thermal ($\nu_x$) or charged current ($\nu_e$ and $\overline{\nu}_e$) processes. As usual, $\nu_x$ is used to represent $\nu_\mu, \nu_\tau, \overline{\nu}_\mu$ and $\overline{\nu}_\tau$ as they to a good approximation behave similarly due to the absence of muons and taus inside a supernova.
\item
Cooling: After the explosion sets in, the accretion stops. However, the protoneutron star continues to emit all types of neutrinos while getting rid of its remaining gravitational binding energy.
\end{enumerate}
See, e.g.,~\citep{jankaneutrinos} for a detailed review on the neutrino emission properties.

\subsection{Neutrino Spectra}
The neutrino spectrum can be well described by a normalized gamma distribution function~\citep{Tamborra:2014hga, Keil:2002in}\\
\begin{align}
f_\alpha=\left(\frac{\alpha+1}{\braket{E}}\right)^{\alpha+1}\frac{E^\alpha}{\Gamma(\alpha+1)}\exp\left(-\frac{(\alpha+1)E}{\braket{E}}\right)\label{spectrum}
\end{align}
with the pinching parameter 
\begin{align}
\alpha=\frac{2\braket{E}^2-\braket{E^2}}{\braket{E^2}-\braket{E}^2}\label{pinching}
\end{align}and the mean neutrino energy $\braket{E}$. Typically, the pinching parameter is in the range of $2<\alpha<3$ except for the initial $\nu_e$ burst which has a stronger pinching of $\alpha\sim6$~\citep{jankaneutrinos}.\\
We will mostly focus on the first $100\,\text{ms}$ of the signal i.e., the initial burst and the rise of the signal during the accretion phase. We used a set of $18$ spherically symmetric SN simulations from~\citep{huedepohl_thesis} and~\citep{risetimemasshierarchy} based on the Lattimer \& Swesty equation of state (EOS) with a bulk incompressibility of $180\,$MeV and $220\,$MeV. However, the different choices of this parameter do not influence the neutrino signal that we are interested in. The models span from SN progenitors with $11.2M_\odot$ up to $40.0M_\odot$. Multidimensional effects seem to not change the general shape during the first $100\,$ms post bounce as, e.g.,~\citep{Tamborra:2014hga} and~\citep{Marek:2008qi} indicate. Also the influence of different EOS on the early signal is very small~\citep{Marek:2008qi,Pan:2017tpk} and should therefore not significantly change our results for timing the onset of the burst. For the black hole (BH) forming case, however, the EOS could have a significant impact. Besides influencing the BH formation probability, a stiffer EOS can shift the formation time further away from core bounce~\citep{Pan:2017tpk}. This would result in a change in the event rate for late time collapses and thereby impact our results.

\section{Detection}\label{sec3}
We focused on three different types of detectors: a liquid scintillator detector (JUNO)~\citep{Djurcic:2015vqa}, two Water-Cherenkov detectors Super-Kamiokande (SK) and Hyper-Kamiokande (HK)~\citep{Abe:2018uyc} and IceCube (IC)~\citep{doi:10.1063/1.3480478}. 
JUNO will be a liquid scintillator detector filled with $20\,\text{kton}$ of linear alkylbenzene (LAB).
The expected threshold for $e^\pm$ detection is
\begin{align}
T_{\text{e, JUNO}}^\text{min}=0.2\,\text{MeV}
\end{align}
and due to quenching
\begin{align}
T_{\text{p, JUNO}}^\text{min}\simeq1\,\text{MeV}
\end{align}
for proton detection~\citep{Beacom:2002hs}. Note that the detection of elastic scattering on protons is very sensitive to this value.\\
Super-Kamiokande is a Water-Cherenkov detector with a fiducial mass of $22.5\,\text{kton}$ and a threshold for electron and positron energies of~\citep{Abe:2016nxk}
\begin{align}
T_{\text{e, SK/HK}}^\text{min}=7\,\text{MeV}\;\;\;.
\end{align}
In the case of a Galactic supernova, however, the detection rate will be much higher than the background so that it will be possible to use the full inner detector volume of $32\,\text{kton}$ for detection. Therefore following the Hyper-Kamiokande design report~\citep{Abe:2018uyc}, we use the entire inner detector mass of $32\,\text{kton}$ for SK and presumably $220\,\text{kton}$ for HK in our calculations. Although HK is expected to have a lower electron kinetic energy threshold of $\sim 3\,$MeV, we keep the more conservative SK threshold for HK too. Note that the event rates after core bounce in SK and HK are not very sensitive to the chosen threshold, since in the major inverse beta decay (IBD) channel most events are above 10MeV (see FIG.\ref{Energyevolution}).\\
Unlike low background detectors with a high photomultiplier tube coverage like SK/HK and JUNO, IceCube will not be able to detect single SN neutrino events. Instead, IC will see a simultaneous increase in Cherenkov light in all of its digital optical modules (DOMs). We calculated the IceCube SN neutrino signal only including the IBD channel~\citep{2009arXiv0908.0441K} both for IC with 5160 DOMs and for a future IC Gen2 with additional 9600 DOMs with a $25\%$ increased dark noise~\citep{ICupgrade, Aartsen:2014njl}.

\subsection{Calculating Event Rates}
In general, the event rate for a certain interaction process $x$ can be calculated from the differential cross section $\frac{\partial\sigma}{\partial T}$ 
and the spectral flux given in terms of the flavor dependent luminosity $L_\nu$ with $\nu\in(\nu_e,\overline{\nu}_e, \nu_x)$ and the distance $D$ as
\begin{align}
F_{\alpha,\nu}=F_\nu f_{\alpha,\nu}=\frac{1}{4\pi D^2}\frac{L_\nu}{\braket{E_\nu}} f_{\alpha,\nu}
\end{align}
via
\begin{align}
R_x=N_\text{t}\sum_\nu\int_{E_{\text{$\nu$, min}}}^\infty \hspace{-4mm} dE_\nu\int_{T_{\text{t, min}}}^{T_{\text{t, max}}(E_\nu)}F_{\alpha,\nu}(E_\nu)\frac{\partial\sigma_x}{\partial T_\text{t}}dT_\text{t}\;\;\;.
\end{align}
Here, $N_{\text{t}}$ is the number of target particles, and $T_{\text{t}}$ is the kinetic energy of the directly detected particle. The sum $\sum_\nu$ runs over all neutrino flavors relevant for the considered interaction. $T_{\text{t, min}}$ is given by the detector threshold and $E_{\text{$\nu$, min}}$ is the corresponding minimal neutrino energy. The mean energy of the detected neutrinos is given by
\begin{align}
\braket{E_{\nu, \text{det}}}=\frac{1}{R}\sum_{x,\nu}\int_{E_{\text{$\nu$, min}}}^\infty \hspace{-4mm} dE_\nu E_\nu\int_{T_{\text{t, min}}}^{T_{\text{t, max}}(E_\nu)}F_{\alpha, \nu}\frac{\partial\sigma_x}{\partial T_\text{t}}dT_\text{t}\;\;\;,
\end{align}
where $R=\sum_x R_x$. (The evolution of the mean energy of the detected neutrinos obtained from this is shown in FIG.~\ref{Energyevolution} taking the ls180s12.0 model as a reference.)\\
To calculate the expected signal for each detector, we included up to three different detection channels, namely the IBD, neutrino-electron elastic scattering, and, in the case of JUNO, also neutrino-proton elastic scattering which accounts for a significant fraction of the overall signal in JUNO due to its lower threshold. 
Also, we assume the SN to be at a distance of $10\,$kpc.\\
In the case of IBD, the recoil energy of the proton can be ignored so that the energy of the detected positron is given by
\begin{align}
E_{e^+}=E_\nu - \Delta m_{np}\quad,\qquad\Delta m_{np}=1.293\,\text{MeV}\label{positron_energy}
\end{align}
For IBD we implement a low energy approximation of the cross section valid for $E_\nu < 300\,\text{MeV}$~\citep{Strumia:2003zx} 
\begin{align}
\sigma_{\text{ibd}}=&10^{-43}\,\text{cm}^2T_eE_eE_\nu^{K(E_\nu)}
\end{align} 
with
\begin{align}
K(E_\nu)=-0.07056+0.02018\log{E_\nu}-0.001953(\log{E_\nu})^3 .
\end{align}
The differential cross section for neutrino-electron elastic scattering at tree level is given by~\citep{Marciano:2003eq,giunti}
\begin{align}
\frac{\partial\sigma_{\nu, e}}{\partial T_e} = \frac{\sigma_0}{m_e}\left[(g_1^{\nu})^2+(g_2^\nu)^2\left(1-\frac{T_e}{E_\nu}\right)^2-g_1^\nu g_2^\nu\frac{m_eT_e}{E_\nu^2}\right]\label{dsigma/dTe}
\end{align}
with
\begin{align}
g_1^{\nu_e}=g_2^{\overline{\nu}_e}=\sin^2\theta_W+\frac{1}{2}\approx0.73&\;,\\
g_2^{\nu_e}=g_1^{\overline{\nu}_e}=g_2^{\nu_x}=g_1^{\overline{\nu}_x}=\sin^2\theta_W\approx0.23&\;,\\
g_1^{\nu_x}=g_2^{\overline{\nu}_x}=\sin^2\theta_W-\frac{1}{2}\approx-0.27&\;,
\end{align}
and
\begin{align}
\sigma_0=\frac{2G_F^2m_e^2}{\pi}\;\;\;.
\end{align}
For the neutrino-proton elastic scattering, we implement 
the differential cross section~\citep{Beacom:2002hs}
\begin{align}
\begin{split}
\frac{\partial\sigma}{\partial T_p}=\frac{\sigma_0m_p}{4m_e^2E_\nu^2}&\left[\left(g_V+g_A\right)^2E_\nu^2\right.
\\&+\left(g_V-g_A\right)^2\left(E_\nu-T_p\right)^2
\\& \left.-\left(g_V^2-g_A^2\right)m_pT_p\right]
\end{split}
\end{align}
with the neutral current axial and vector couplings
\begin{gather}
g_A=\frac{1.27}{2}\;,\\
g_V=\frac{1-4\sin^2\theta_W}{2}\;.
\end{gather}

\begin{figure*}[tbp]
\includegraphics[width=0.75\textwidth]{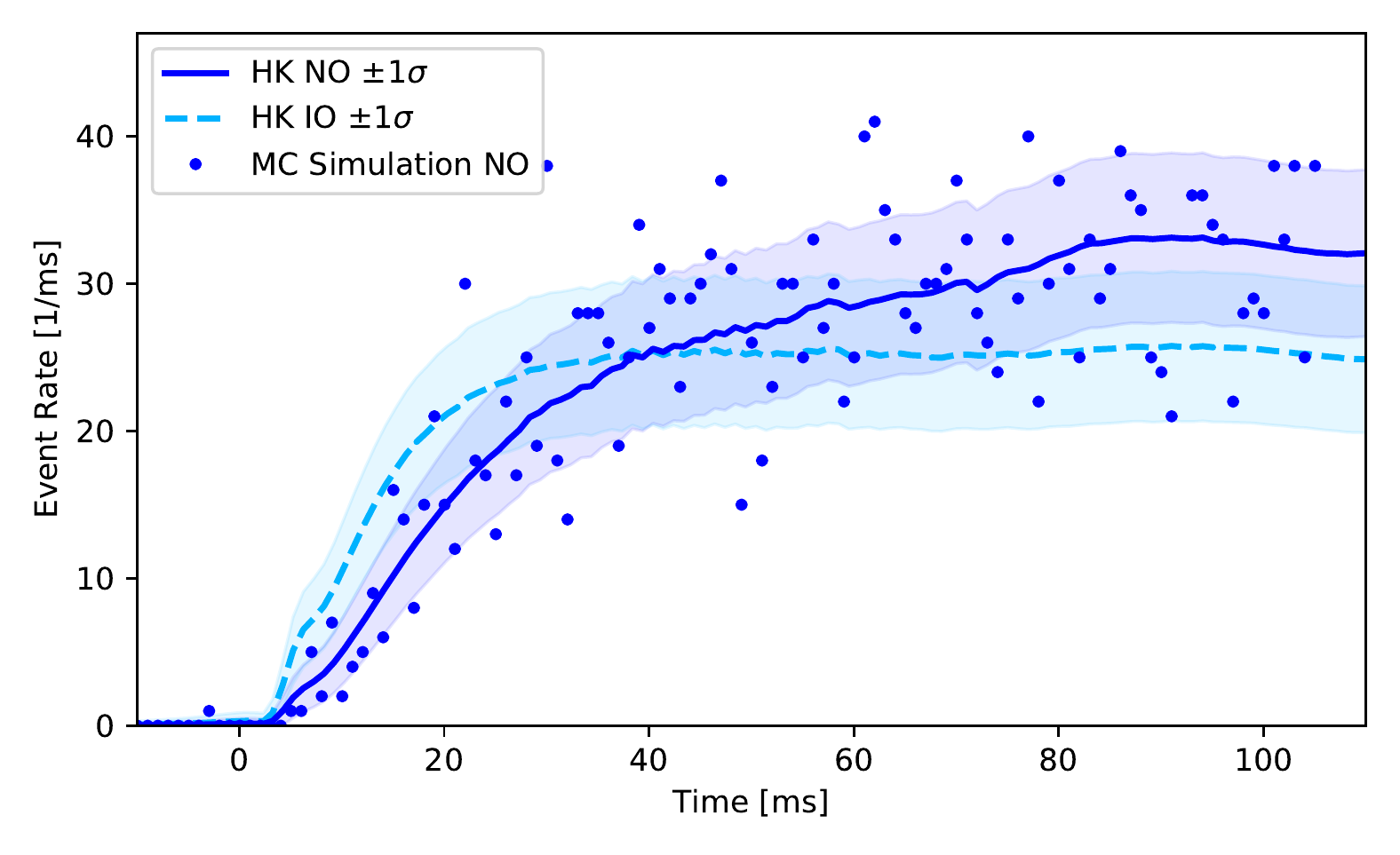}\\
\includegraphics[width=0.49\textwidth]{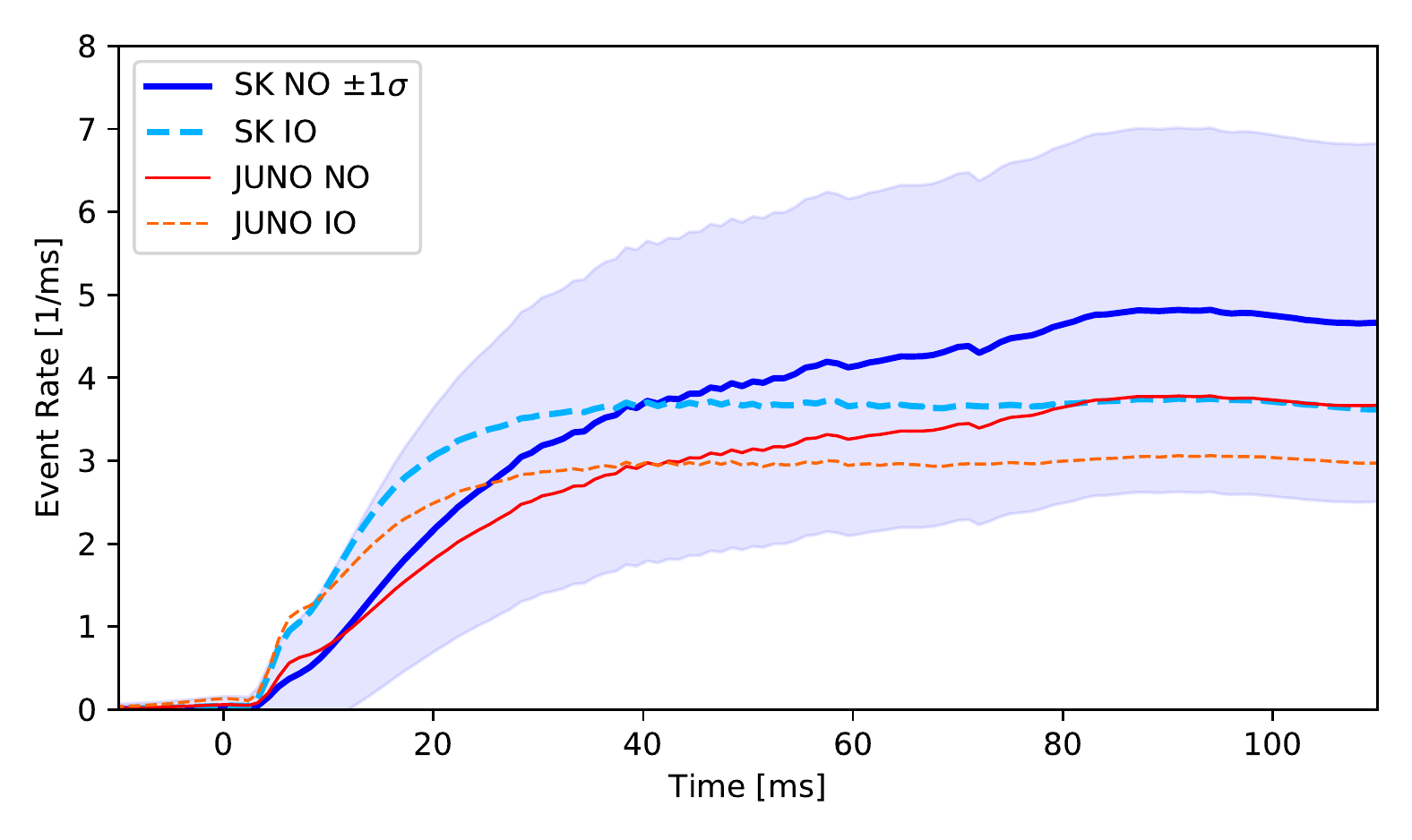}
\includegraphics[width=0.49\textwidth]{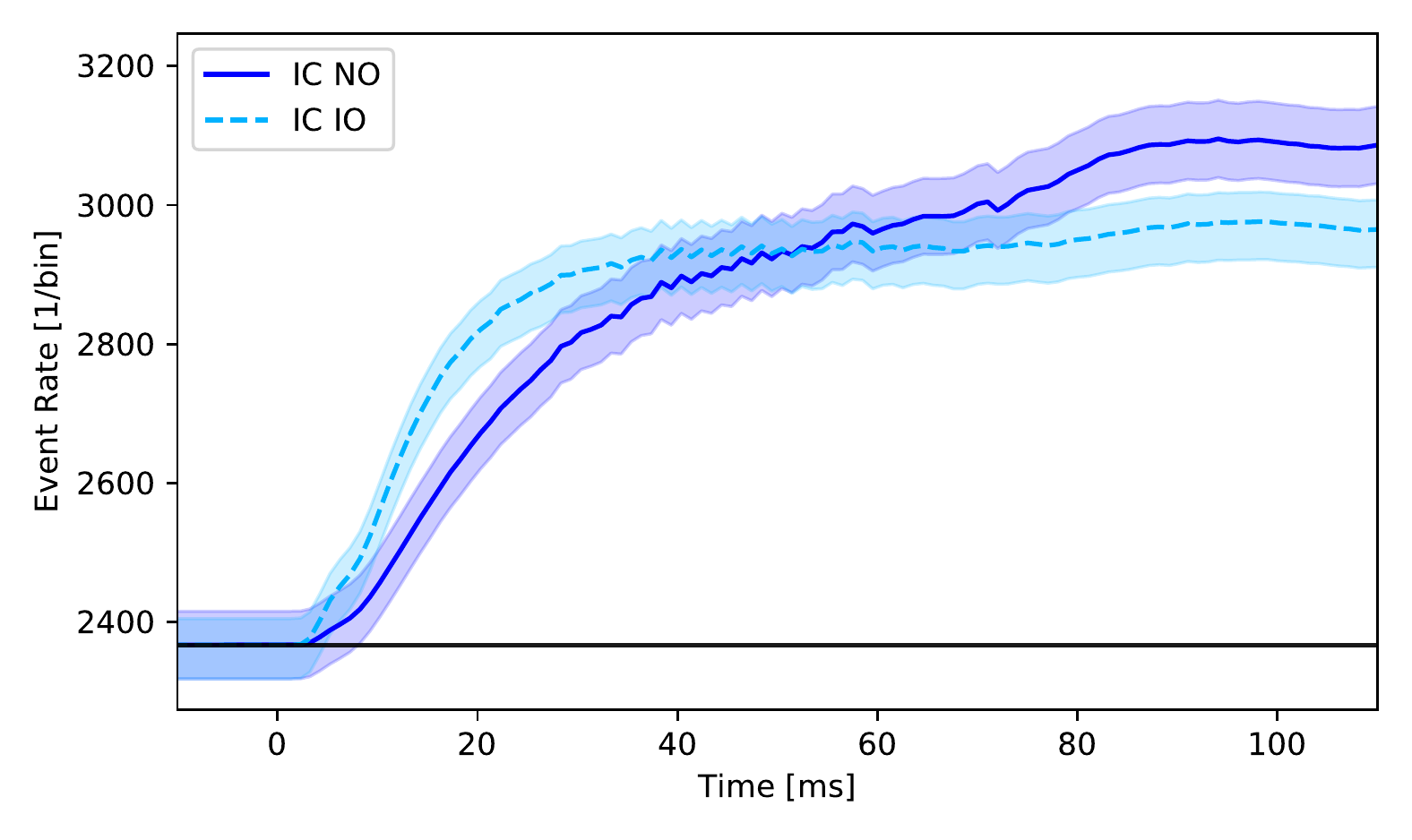}
\caption{Mean eventrates in HK (upper), SK, JUNO (left) and IC (right) using the ls180s12.0 simulation from Hüdepohl~\citep{risetimemasshierarchy}. The shaded areas show the $1\sigma$ deviation. The blue dots in the upper panel show one MC realization in HK assuming normal ordering. Note that the rates for IC are given per bin i.e. per $1.6384\,\text{ms}$. The black horizontal line in the lower right panel represents the constant background noise of $280\,$Hz per module in IC. A sample MC realization for IO as well as more details on the timing methods are displayed in FIG.~\ref{expfitplot}.}\label{burstrates}
\end{figure*}

\subsection{Neutrino Flavor Conversion}
To convert the individual fluxes and spectra of each neutrino flavor at the supernova to the observed signal at Earth, neutrino flavor conversion must be taken into account. A recent study suggests that in the case of a failed supernova, collective oscillations can be ignored~\citep{Zaizen:2018wfg} such that only matter effects need to be considered. In general, however, collective effects could play an important role in determining the final fluxes \citep{Mirizzi:2015eza, Duan:2010bg}. In the following we will only consider adiabatic Mikheyev-Smirnov-Wolfenstein (MSW) conversion. Also we assume the SN and the detectors to be within the same hemisphere so that we can ignore any Earth-matter effects.\\
In the high density neutrinosphere, the Hamiltonian becomes effectively diagonal in flavor space such that pure Hamiltonian eigenstates $\nu_{1m}, \nu_{2m}$ and $\nu_{3m}$ are produced~\citep{Dighe:1999bi}. Those propagate outwards through the SN and are converted to the vacuum eigenstates $\nu_1, \nu_2$ and $\nu_3$. Depending on the mass hierarchy, one finds the final fluxes of the different neutrino mass eigenstates in terms of the initial flavor fluxes $F^0_{\alpha}$ as it is shown in TABLE~\ref{msw}.
\begin{table}[tbp]
\begin{tabularx}{\columnwidth}{Y Y Y Y }
\hline\hline
\multicolumn{2}{c }{Normal ordering (NO)} & \multicolumn{2}{c}{Inverted ordering (IO)} \\ 
 $\nu$ & $\overline{\nu}$ & $\nu$ & $\overline{\nu}$  \\
 \hline
$F_{\nu_1}=F_{\nu_x}^0$ & $F_{\overline{\nu}_1}=F_{\overline{\nu}_e}^0$ & $F_{\nu_1}=F_{\nu_x}^0$ &$F_{\overline{\nu}_1}=F_{\nu_x}^0$ \\ 
$F_{\nu_2}=F_{\nu_x}^0$ & $F_{\overline{\nu}_2}=F_{\nu_x}^0$ & $F_{\nu_2}=F_{\nu_e}^0$ &$F_{\overline{\nu}_2}=F_{\nu_x}^0$ \\ 
$F_{\nu_3}=F_{\nu_e}^0$ & $F_{\overline{\nu}_3}=F_{\nu_x}^0$ & $F_{\nu_3}=F_{\nu_x}^0$ &$F_{\overline{\nu}_3}=F_{\overline{\nu}_e}^0$ \\ 
\hline\hline
\end{tabularx} 
\caption{{Initial $F^0_{\nu_\alpha}$ and final $F_{\nu_i}$ total fluxes for neutrinos $\nu$ and anti-neutrinos $\overline{\nu}$ depending on the mass hierarchy.}}\label{msw}
\end{table}
Consequently, the final fluxes of the different flavors at Earth are given by
\begin{align}
F_{\nu_\alpha}=\sum_i |U_{\alpha i}|^2 F_{\nu_i}
\end{align}
with $U$ being the Pontecorvo–Maki–Nakagawa–Sakata (PMNS) matrix. Correspondingly, the final normalized spectra are given by 
\begin{align}
f_{\nu_\alpha}=\frac{\sum_i |U_{\alpha i}|^2 f_i^0 F_{\nu_i}}{\sum_i F_{\nu_i}} \;.
\end{align}
Since the difference in the mean energies of the different neutrino mass eigenstates are small compared to the width of the initial spectra, these final neutrino spectra for the different flavors at Earth are also described well by a gammalike distribution. Note that this only works if the initial spectra overlap strongly.

\subsection{Background}
\label{sec:background}
Compared to the event rate during a SN, the background in the relevant energy range is many orders of magnitude smaller. In JUNO  0.001 background events are expected per second~\citep{An_2016} while 0.01 per second are expected for SK~\citep{, Abe:2016waf}.
Therefore these backgrounds are negligible in JUNO as well as in SK when being compared to the enormous SN rate. Treating HK as an up-scaled SK, we expect $\sim 0.1$ background events per second in HK~\citep{Abe:2018uyc}. Since we are considering intervals of 100ms or less, we did not include any background events in our MC simulations for those three detectors. In the case of IceCube however, a SN will be detected by noise excess rather than a measurement of single events. For IC the background can be treated as a Poissonian background with an expectation value of $280\,$Hz per module ($+25\%$ expected for the IC Gen2 modules)~\citep{2009arXiv0908.0441K, ICupgrade, Aartsen:2014njl}.

\subsection{Event Rates in the Detectors}
The expected event rates calculated by the above means using the ls180s12.0 SN simulation from Hüdepohl~\citep{risetimemasshierarchy} are shown for each detector and both normal and inverted ordering in FIG.~\ref{burstrates}.
By comparing the three panels in FIG.~\ref{burstrates}, we clearly see the benefit of a larger detector as the $1\sigma$ error bands shrink as we go from JUNO/SK to HK and to IC. On the other hand, the most promising feature for timing the signal is the onset, and here IC suffers from the high background rate, which means that the fluctuation in the number of events at time zero is comparatively large.
Another feature that is worth noticing is the difference in the rise of the signal between NO and IO. As pointed out in~\citep{risetimemasshierarchy}, this can be used to pinpoint the neutrino mass ordering in the event of a reasonably nearby Galactic supernova.\\
Additonally, FIG.~\ref{scatrates} shows the expected rate from elastic scattering on electrons in HK separately which will be discussed in Sec.~\ref{sec4b} in more detail.

\section{Timing the Signal}\label{sec4}
We have investigated several methods using characteristic structures of the neutrino signal for timing purposes.
This is done with a Monte Carlo (MC) simulation of the neutrino signal with $10000$ realizations in each detector for each of the $18$ SN
simulations available to us. Each MC simulation was done both for NO and for IO.
For each Monte Carlo run, we use the average detector rates calculated and assume a Poissonian distribution. The total time period of the signal is binned into $1\,\mu$s bins to produce "single" neutrino events. 
One of the MC realizations for HK is shown in FIG.~\ref{burstrates} for comparison with the mean rates. It is clear that the NO and IO will be different despite the random scattering of the events. We also notice that there are a few events before the time of the burst.
In FIG.~\ref{expfitplot} we show a visualization for each of the timing methods presented below for a single Monte Carlo realization in Hyper-Kamiokande for both normal and inverted ordering.\\
In the following, we present the investigated methods. The averaged results are summarized in TABLE~\ref{avrg_results}. It shows the standard deviation of the timing variable for each method presented below averaged over all SN models. The average is taken over the uncertainties as the absolute timing in many cases is expected either to be of little importance (e.g. when using the timing for triangulation with identical detectors) or to be determined by detailed modeling (e.g. for doing a time of flight measurement relative to gravitational waves where the two signals need to be related through a model in the first place). The absolute timings as well as the standard deviation for each single MC simulation can be found in Appendix~\ref{App. A} while results assuming different non-zero neutrino masses are presented in Appendix~\ref{App. B}.

\subsection{Exponential Fit}\label{sec4a}
\begin{figure*}[tbp]
\includegraphics[width=\columnwidth]{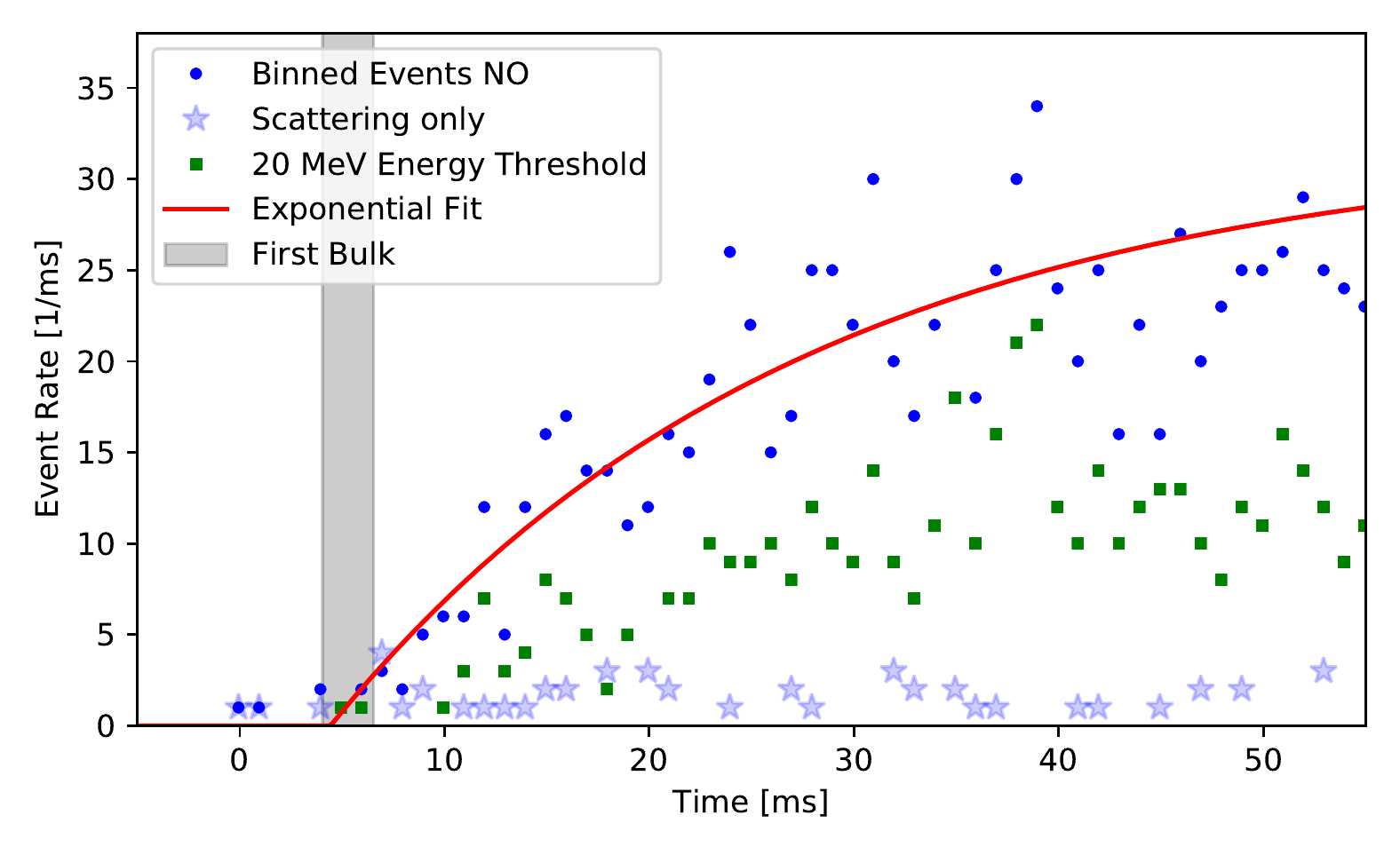}
\includegraphics[width=\columnwidth]{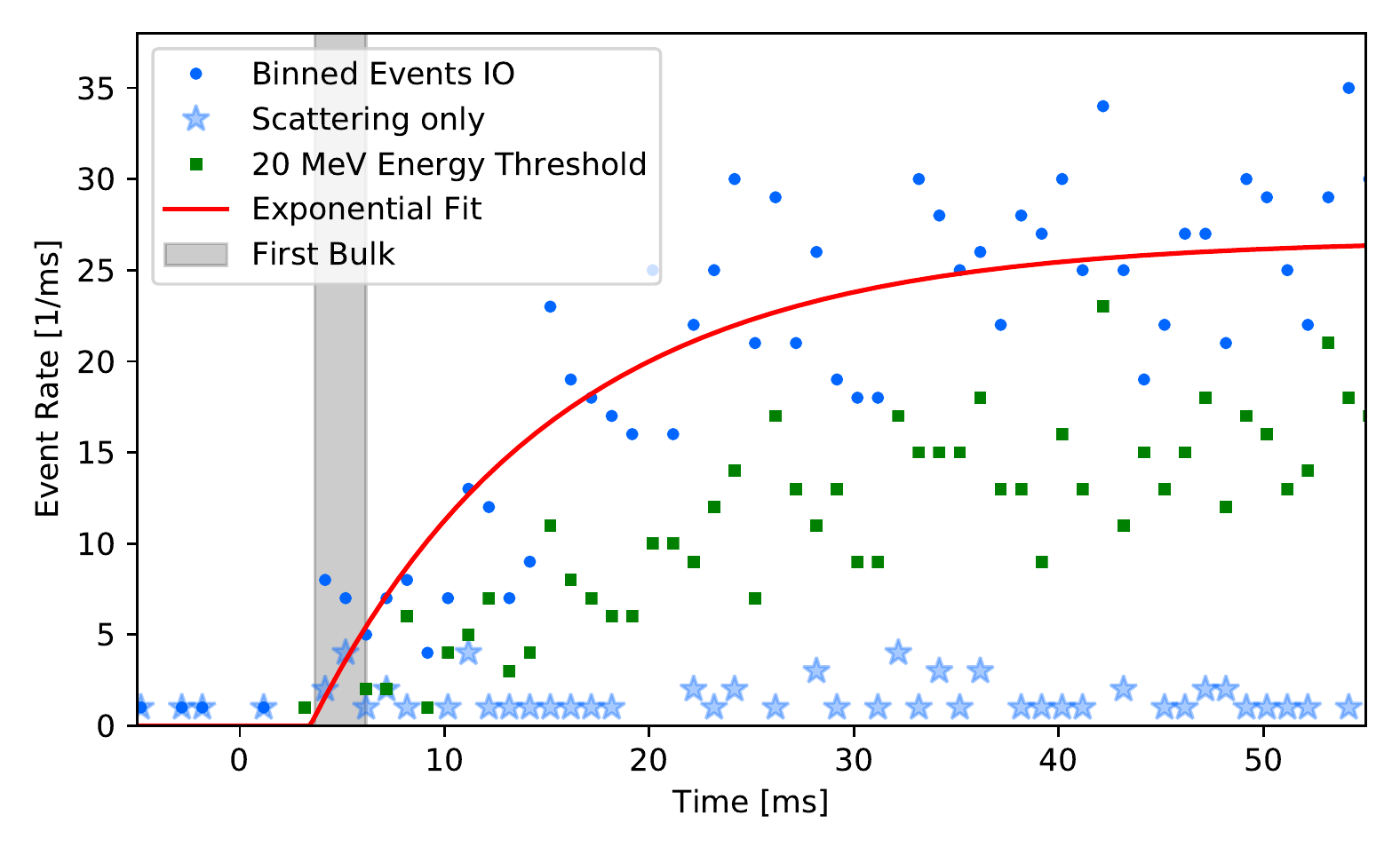}
\caption{{Visualization of the different methods in Sec.~\ref{sec4} for HK (Exponential Fit, First Bulk, Energy Threshold and First IBD) for NO (left panel) and IO (right panel). The (light-)blue dots represent the total binned signal of one specific Monte Carlo realization, the red curve shows the fit resulting from Eq. \eqref{expfitequation}, the green squares show the binned events that produce secondary $e^\pm$  with a kinetic energy $T_e>20\,$MeV, and the (light-)blue stars show the binned elastic scattering event rate. Therefore, the first green square shows the bin with the event that triggers the Energy Threshold method, while the first bin in which the blue dot and star do not match shows the bin in which the first IBD event is located. The gray area shows the $2.5\,$ms time period of the first bulk that was found. Note that the timing of the single events is taken to be the actual time of the event and not the time of the corresponding bin, thus obtaining sub-ms resolution.}}\label{expfitplot}
\end{figure*}

A method for timing the supernova neutrino burst which was already explored for IC~\citep{PhysRevD.80.087301} is fitting a function of the form
\begin{align}
R_{\text{exp}} = R_{\text{max}}\cdot\begin{cases} 0 & t<t_0\\ \left(1-\exp{\left[(t-t_0)/\tau\right]}\right) & t>t_0 \end{cases}\label{expfitequation}
\end{align}
to the measured rate, where $t_0$ represents the onset of the signal which we want to determine ($t_0$ is the timing variable). We further explore this possibility for SK, HK, and JUNO detectors as well as a potential IceCube Gen2 upgrade by fitting the first $100\,$ms of the signal.\\
To fit the rate in SK, HK, and JUNO, the signal was rebinned into $1$ ms bins. The results depend only weakly on this choice, but tend to be slightly worse for much larger bins.
The two fits in FIG.~\ref{expfitplot} show how $R_{\text{exp}}$ rises and then goes to a plateau. For NO in the left panel, the plateau is only reached at times larger than 50ms, but for IO in the right panel, the plateau starts around 30ms. The uncertainty in timing the onset of the signal is just above a ms for NO and just below for IO in the case of HK.

\begin{figure}[tbp]
\includegraphics[width=\columnwidth]{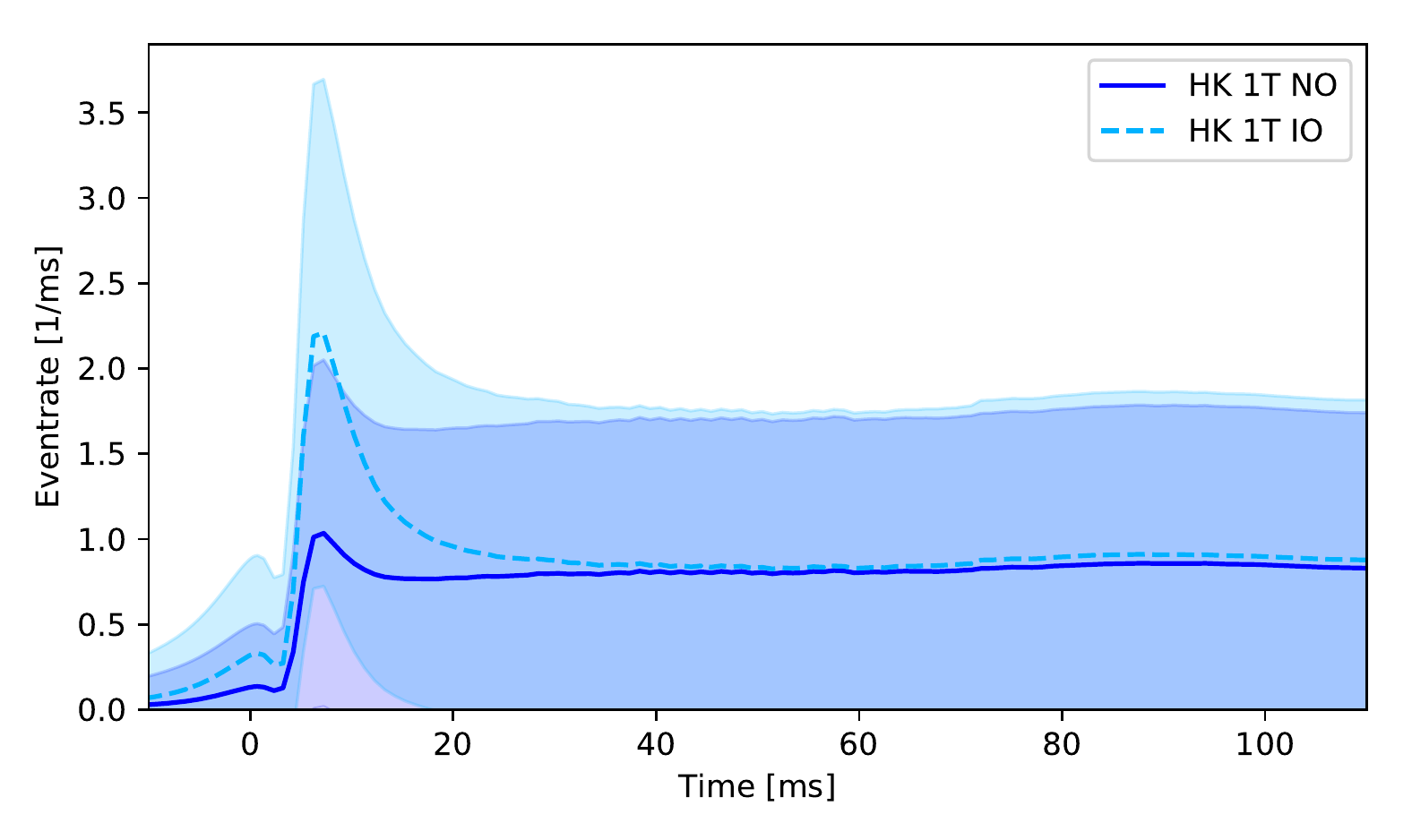}\caption{Mean elastic neutrino-electron scattering rates $\pm1\sigma$ in HK for NO (solid line) and IO (dashed line).
}\label{scatrates}
\end{figure}

\subsection{Gauss Fit of the Initial $\nu_e$-Burst}\label{sec4b}
Having a look at the first part of FIG.~\ref{emissionphases}, the characteristic structure of the strong $\nu_e$-burst seems to be a promising candidate for a timing reference. However, looking at the early times of the simulated signals in FIG.~\ref{burstrates}, it only leads to a very small bump in the signal.
This has mainly two reasons:
\begin{enumerate}
\item The cross section for elastic scattering on electrons is much smaller than the cross section for IBD.
\item The cross section for elastic scattering on electrons is
higher for $\nu_e$ and $\overline{\nu}_e$ than for other flavors since both NC 
and CC elastic scattering can occur while it is larger for $\nu_e$ than for $\overline{\nu}_e$ due to the different helicities of neutrinos and antineutrinos. Because of matter effects, the initial $\nu_e$ flux $F_{\nu_e}^0$ corresponds to $F_{\nu_3}$ in case of NO and $F_{\nu_2}$ in case of IO (see TABLE~\ref{msw}). Since the initial $\nu$-burst consists almost only of $\nu_e$, the final $\nu_e$ flux at the detector is suppressed by the smallness of $|U_{e3}|^2\approx0.02$ (NO) or $|U_{e2}|^2\approx0.3$ (IO).
\end{enumerate}
The small bump in the expected detector rate resulting from elastic scattering events will therefore not be visible in the total signal since it will be dominated by the Poissonian fluctuations. JUNO, however, will be able to distinguish elastic scattering and IBD events via neutron capture~\citep{Djurcic:2015vqa}. The same might be achieved in SK/HK with the use of gadolinium~\citep{2009arXiv0909.5528K}.\\
Assuming a (rather optimistic) perfect identification of IBD vs. elastic scattering events, we further explored the possibility of detecting and timing the peak by fitting
\begin{align}
R = 
\begin{cases} 
0 & t<t_0\\ 
a\cdot\exp\left[\frac{-(t-t_\text{burst})^2}{b}\right]+R_{\text{exp}}
& t>t_0 
\end{cases}
\end{align}
to the first 100ms and determining $t_\text{burst}$ as the timing variable.
Taking a look at FIG.~\ref{scatrates} for HK, one would expect to see the peak in some of the MC realizations in the case of IO since it differs from the following plateau by a little more than $1\,\sigma$ ($\sqrt{R}$), while one would expect to see no peak in most of the MC realizations for NO. To prevent overfitting we only took into account fits with a peak full width at half maximum (FWHM) of $30\,\text{ms}>\text{FWHM}>3\,\text{ms}$. Our MC simulations (see Gauss Fit in TABLE~\ref{avrg_results}) show that the fits for SK and JUNO have high fail rates, so there is little to no chance to see the peak as expected, while for HK it might be possible under the given assumptions in some cases. As we mentioned, however, the assumption of a perfect identification is rather optimistic, and in reality, the IBD identification efficiency in a gadolinium filled water Cherenkov detector the size of SK will be $\sim 50-90\%$~\citep{Labarga:2018fgu} depending on the amount of gadolinium.
In HK, the expected efficiency is $73\%$ without and up to $90\%$ with gadolinium~\cite{Abe:2018uyc}.

\subsection{Identifying the First Neutrino After Core Bounce}
Detectors such as the Kamiokande detectors or JUNO provide the unique opportunity to identify the timing of single neutrino events, therefore eliminating statistical errors that may arise from the above fitting methods.\\
Looking at the two MC realizations in FIG.~\ref{expfitplot}, however, it is clear that the first neutrino that is detected, was emitted prior to the neutrino burst itself. Hence a method to exclude preburst neutrinos is necessary.

\subsection*{First Bulk}
After core bounce, the neutrino event rate increases rapidly. It is therefore quite natural to define the first "bulk"
of neutrino events as the start of the supernova neutrino burst. To define this bulk more quantitatively, we can use the exponential fit from Sec.~\ref{sec4a} and integrate it over the first $2.5$ ms in HK and $20$ ms in SK/JUNO. Then we can search for the earliest neutrino event which is inside such a bulk with $N>N_{\text{integrated}}$ neutrino events and take the time of that event as our timing variable. The results only depend weakly on the integration time as long as it is sufficiently large to catch several events.
In FIG.~\ref{expfitplot}, the First Bulk is marked by the grey region. For the example with NO, there is a slight offset to earlier times, while for IO it is to later times when compared to the onset of the Exponential Fit. This is the typical behaviour for HK. In SK and JUNO due to the smaller event rate, the First Bulk timing is offset to later times in both NO and IO. The timing uncertainty is on average slightly smaller than for the Exponential Fit method. We attribute this to the sharp rise in the event rate which is captured by the First Bulk method, without being affected by random fluctuations at later times which the Exponential Fit is more prone to.

\subsection*{Energy Threshold}
\begin{figure}[tbp]
\includegraphics[width=\columnwidth]{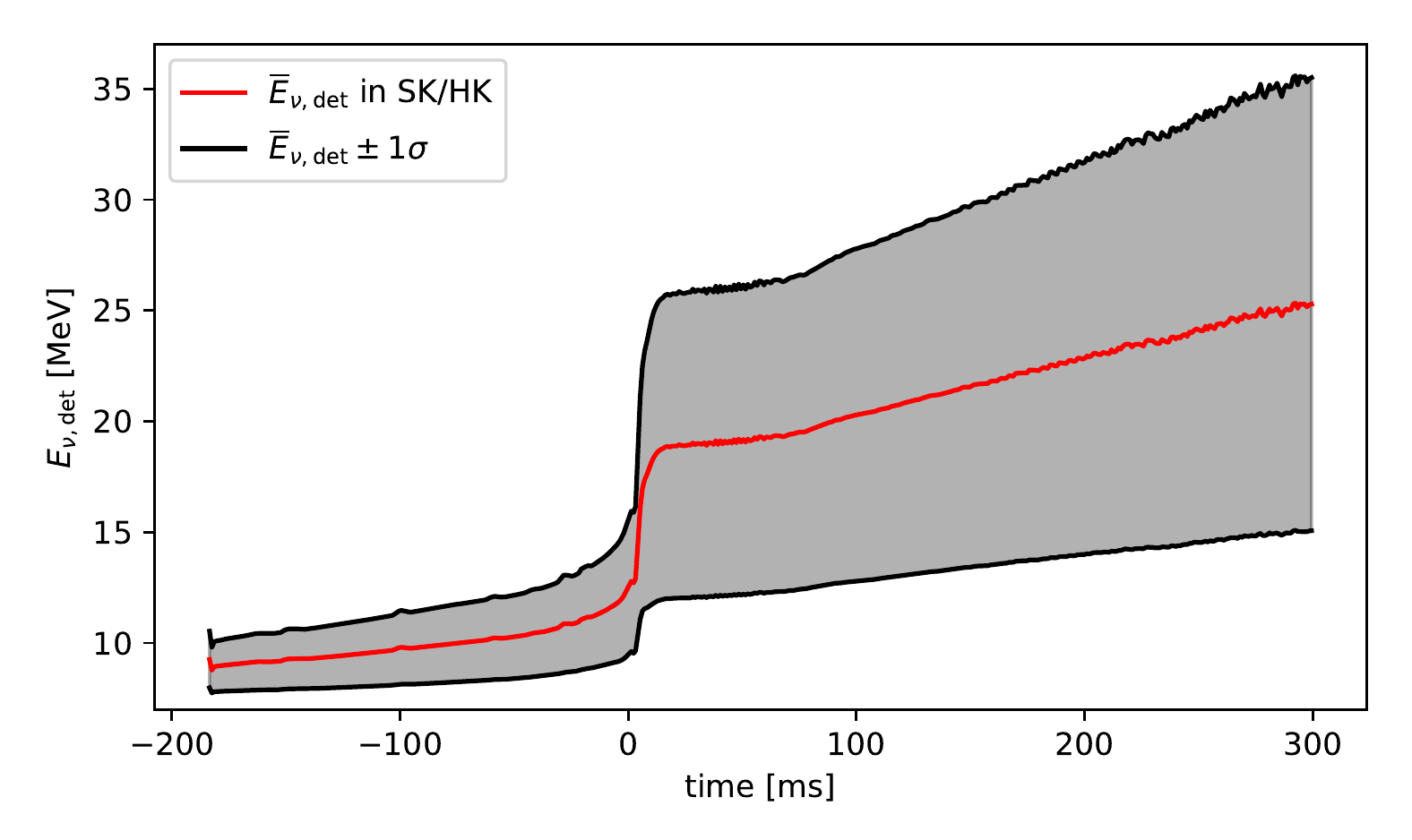}
\caption{{Evolution of the mean energy of the detected neutrinos in SK/HK for the ls180s12.0 star. The features are found to be independent of the progenitors and the EOS of stars in the dataset used.}}\label{Energyevolution}
\end{figure}

Another approach to distinguish preburst neutrinos from postburst neutrinos is to look at the energies of the single events that are detected. Looking at FIG.~\ref{Energyevolution}, one can see that there is a sudden increase in the mean energy of the detected neutrinos at the core bounce. In a detector, we do not measure the neutrino energy directly, but rather the energy of the secondary particle ($e^-, e^+ , p$ ...) which will be lower. Still, it is possible to define the first postbounce neutrino as the first event with a secondary particle energy 
above $E_{\rm threshold}$ which we put to $20$ MeV for most of our analysis. The time of this event is our timing variable for the Energy Threshold method.
In our Monte Carlo realizations, the energy of the scattered electrons was simulated assuming that the spectrum of the detected neutrinos 
follows the gamma distribution of Eq. \eqref{spectrum} with the pinching parameter fixed by $\braket{E}$ and $\braket{E^2}$ according to Eq. \eqref{pinching}. This assumption is reasonable since the spectra of the different neutrino flavors are quite similar, and the detector thresholds are well below the mean energy $\braket{E}$ 
during the relevant time after core bounce.
Since we are only interested in events above $20\,$MeV, we can define the spectral difference as
\begin{align}
\Delta f = \frac{\int_{E_{\rm threshold}}^\infty dE\,\left[f_{\text{gamma}}(t,E)-f_{\text{real}}(t,E)\right]}{\int_{E_{\rm threshold}}^\infty dE\,\left[f_{\text{gamma}}(t,E)+f_{\text{real}}(t,E)\right]}
\end{align}
where $f_{\rm gamma}$ is a gammalike spectrum as in Eq.~\eqref{spectrum} given by the mean and squared mean electron recoil energy while $f_{\rm real}$ is the exact spectrum given the full neutrino spectrum from the SN simulations.
During the relevant time this difference is $1\%<\Delta f < 2\%$.
In general, this approximation overestimates the spectrum near its peak while underestimating the spectrum for higher energies. The energies of preburst neutrinos, however, are overestimated by this approximation because the mean neutrino energy is still close to the detector thresholds. However, due to the low mean neutrino energy at these early prebounce times, the normalized spectra are close to zero, in the energy range $E>E_{\rm threshold}$ relevant for our analysis i.e.
\begin{align}
\begin{split}
&f_{\rm real}(t<t_{\rm bounce}, E>E_{\rm threshold})\\
<&f_{\rm gamma}(t<t_{\rm bounce}, E>E_{\rm threshold})\sim 0
\end{split}
\end{align}
such that the overestimation is not relevant for us. In other words, although we overestimate the energy of the neutrinos during prebounce time, the spectral shape is such that the probability to detect a neutrino above $E_{\rm threshold}$ is practically zero.
The spectrum of the scattered electrons for a fixed neutrino energy is then given by the differential cross section \eqref{dsigma/dTe}. For IBD events, the energy of the produced positrons is well approximated by \eqref{positron_energy}. For our purpose, we can ignore the scattered protons in liquid scintillator detectors here since their energy is significantly lower than the energy of the secondary particles from IBD and electron elastic scattering due to the higher mass of the proton.\\
The events with energies above the threshold are shown by green squares in FIG.~\ref{expfitplot}. For NO, the first event above threshold is located a few ms after the times of Exponential Fit and First Bulk, while for IO the first event is before the timing of the other two methods. Since the Energy Threshold method have a high chance of rejecting the first few neutrinos after bounce, we expect the general offset to be larger as one can see in Appendix~\ref{App. A}. For SK and JUNO the timing uncertainty of this method is $\sim 0.5\,$ms better than for the Exponential Fit method while for HK it turns out to be only slightly larger.

\begin{table*}[tbp]
\begin{tabularx}{\textwidth}{l c Y Y Y Y Y }

\hline\hline
Method & Ordering &   {HK}   &   {SK}   &   {JUNO}   &   {IC}   &   {IC Gen2}  \\
\hline\\
Exponential Fit [ms] & $\begin{matrix} \text{NO}\\ \text{IO} \end{matrix}$ &$\begin{matrix} 1.2 \\ 0.9 \end{matrix}$ & $\begin{matrix} 3.2 \\ 2.6 \end{matrix}$ & $\begin{matrix} 3.8 \\ 3.0 \end{matrix}$ & $\begin{matrix} 1.1 \\ 0.9 \end{matrix}$ & $\begin{matrix} 0.7 \\ 0.6 \end{matrix}$ \\  \\ 
Gauss Fit [ms] & $\begin{matrix} \text{NO}\\ \text{IO} \end{matrix}$ &$\begin{matrix} 4.5 \\ 2.8 \end{matrix}$ & $\begin{matrix} 6.4 \\ 4.6 \end{matrix}$ & $\begin{matrix} 5.6 \\ 4.2 \end{matrix}$ & $\begin{matrix} - \\ - \end{matrix}$ & $\begin{matrix} - \\ - \end{matrix}$ \\  \\ 
Failed Gauss Fits $\%$ & $\begin{matrix} \text{NO}\\ \text{IO} \end{matrix}$ &$\begin{matrix} 42 \\ 26 \end{matrix}$ & $\begin{matrix} 60 \\ 49 \end{matrix}$ & $\begin{matrix} 45 \\ 44 \end{matrix}$ & $\begin{matrix} - \\ - \end{matrix}$ & $\begin{matrix} - \\ - \end{matrix}$ \\  \\ 
First IBD [ms] & $\begin{matrix} \text{NO}\\ \text{IO} \end{matrix}$ &$\begin{matrix} 0.9 \\ 0.6 \end{matrix}$ & $\begin{matrix} 2.2 \\ 1.4 \end{matrix}$ & $\begin{matrix} 2.6 \\ 1.7 \end{matrix}$ & $\begin{matrix} - \\ - \end{matrix}$ & $\begin{matrix} - \\ - \end{matrix}$ \\  \\ 
First Bulk [ms] & $\begin{matrix} \text{NO}\\ \text{IO} \end{matrix}$ &$\begin{matrix} 1.0 \\ 0.7 \end{matrix}$ & $\begin{matrix} 2.8 \\ 2.2 \end{matrix}$ & $\begin{matrix} 3.2 \\ 2.8 \end{matrix}$ & $\begin{matrix} - \\ - \end{matrix}$ & $\begin{matrix} - \\ - \end{matrix}$ \\  \\ 
20 MeV Energy Threshold [ms] & $\begin{matrix} \text{NO}\\ \text{IO} \end{matrix}$ &$\begin{matrix} 1.3 \\ 0.9 \end{matrix}$ & $\begin{matrix} 2.7 \\ 2.0 \end{matrix}$ & $\begin{matrix} 3.2 \\ 2.4 \end{matrix}$ & $\begin{matrix} - \\ - \end{matrix}$ & $\begin{matrix} - \\ - \end{matrix}$ \\  \\ 
Black Hole Collapse [ms] & $\begin{matrix} \text{NO}\\ \text{IO} \end{matrix}$ &$\begin{matrix} 0.012 \\ 0.017 \end{matrix}$ & $\begin{matrix} 0.08 \\ 0.11 \end{matrix}$ & $\begin{matrix} 0.09 \\ 0.12 \end{matrix}$ & $\begin{matrix} - \\ - \end{matrix}$ & $\begin{matrix} - \\ - \end{matrix}$ \\  \\
\end{tabularx}
\caption{Averaged uncertainties. For each method the $1\sigma$ standard deviation of the timing variable as described in Secs. \ref{sec4a} - \ref{sec4d} averaged over all 18 SN simulations is given. The failed fit percentages correspond to fits that either did not converge or were rejected to prevent overfitting (see Section~\ref{sec4b}).}\label{avrg_results}
\end{table*}
\subsection*{First IBD}
Again assuming perfect identification of IBD events, one can define the timing of the first IBD event as the start of the burst (and our timing variable) since the preburst neutrinos consist only of $\nu_e$ (see FIG.~\ref{emissionphases}). The IBD timing scales with
\begin{align}
    \Delta t \propto \frac{1}{\sqrt{R_{\text{IBD}}}}
\end{align}
to a good approximation, where $R_\text{IBD}$ is the neutrino IBD rate. This is due to the linear increase in the event rate seen in FIG~\ref{burstrates}. Since 
\begin{align}
    R_\text{IBD}\propto\epsilon
\end{align} 
where $\epsilon$ is the neutron tagging efficiency, one can rescale our results to any neutron tagging efficiency by multiplying with $\frac{1}{\sqrt{\epsilon}}$. For an expected efficiency of $73\%$, we estimate the average 1$\sigma$ standard deviation for HK and IO to be $0.6\,{\rm ms}/\sqrt{0.73} = 0.7\,$ms. Also, fluctuations in the delay of the neutron capture will play a role for very nearby SN with significantly higher event rates.\\
For the MC realisations in FIG~\ref{expfitplot}, the First IBD timing is represented by the first time when the blue dot and star do not match. In the left part of the figure, one can see that for NO the first such event is just before the timing of the Exponential Fit, and the same is the case in the right part of the figure for IO. This behaviour is typical for NO, while for IO the larger elastic scattering rate before core bounce shifts the Exponential Fit to earlier times such that the First IBD event is on average shortly after the the timing of the Exponential Fit. For SK and JUNO the first IBD event generally happens a few ms after the timing of the Exponential Fit independently of the mass ordering. This is due to the lower event rate for these two detectors. However, in any case the uncertainty in determining the offset is better than for any of the other methods. As is the case for the First Bulk method, the power of the First IBD method is to take full advantage of the quick rise of the signal without relying on a fit to a larger portion of the signal.

\subsection{Black Hole Collapse}\label{sec4d}
Although the exact fraction is still unknown, it is expected that some CCSNe will collapse to a Black Hole (BH)~\citep{OConnor:2010moj, Horiuchi:2014ska}. Observationally, the fraction of these so called failed supernovae is estimated to be~\citep{Adams:2016hit}
\begin{align}
f_{\text{failed}}=0.14^{+0.33}_{-0.10}
\end{align}
at $90\%$ confidence. Other more theoretical works predict a fraction between $\sim0.1-0.4$~\citep{Suzuki:2017kbh, Zhang:2007nw}.
In the case of BH formation happening while the neutrino signal is still measurably high, the neutrino emission will be cut off abruptly when the neutrinosphere falls inside the horizon of the BH. This characteristic cut-off provides another possibility for timing the neutrino signal.\\
For detectors like SK, HK, or JUNO, it is possible to define the cut-off time (our timing variable) as the time of the last detected neutrino event so that the timing resolution is given by the average time between two events i.e. the inverse detection rate at the time of the cut-off
\begin{align}
\Delta t = \frac{1}{R_{\text{det}}}\;\;\;,
\end{align}
where $R_{\rm det}$ is the rate in the detector.
Looking at the Black Hole Collapse line in TABLE~\ref{avrg_results}, this is $\mathcal{O}(10)\,\mu$s for HK. At such small timescales, the formation process of the BH itself will start to play a significant role. Assuming a protoneutron star radius of $10\,$km, we can estimate the BH formation timescale with the help of the light-crossing time which in this case will be $\sim 70\,\mu$s i.e. more than $4$ times larger than the estimated resolution in Hyper-Kamiokande. Early numerical SN simulations show that the collapsing time for an actual observer at Earth will be $\mathcal{O}(0.5)\,$ms~\citep{1996ApJ468823B}. In the case of a failed SN happening in our galaxy, Hyper-Kamiokande might therefore allow us to observe the process of a protoneutron star collapsing to a black hole. However, this will strongly depend on the real distance $D$ to the failed supernova since the event rate and therefore the timing resolution scales with $D^{-2}$.
\begin{figure*}[t]
\includegraphics[width=0.497\textwidth]{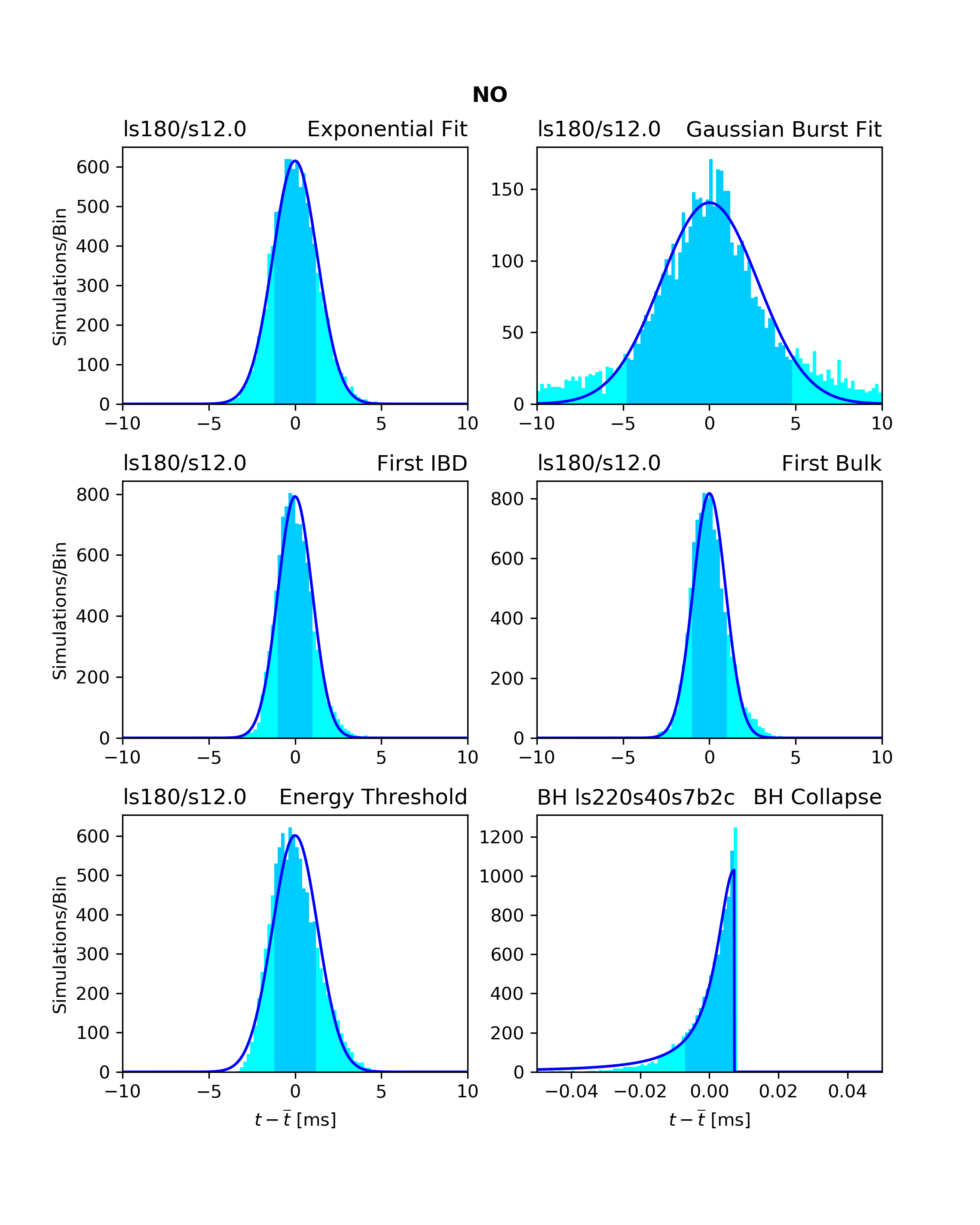}
\includegraphics[width=0.497\textwidth]{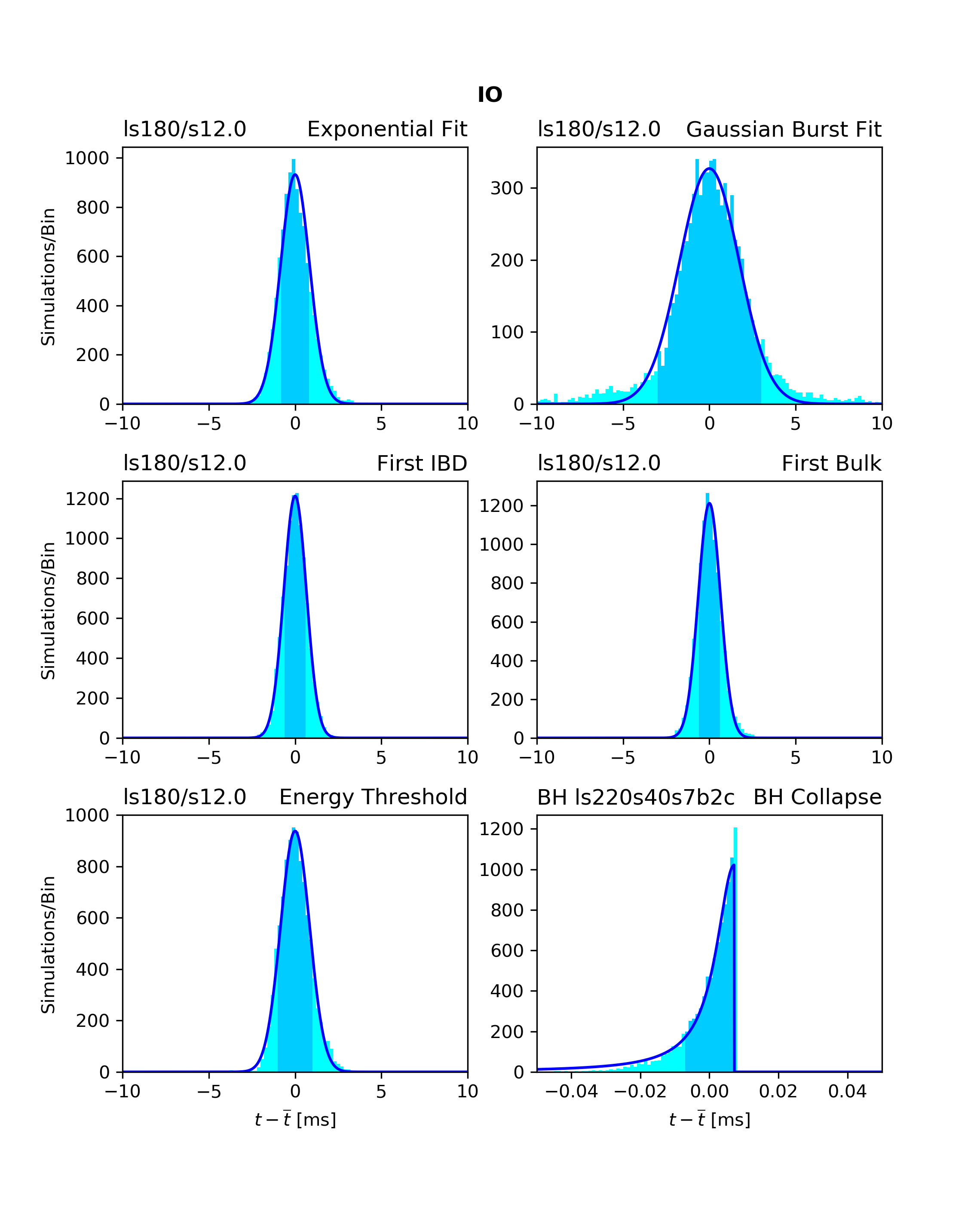}\\
\caption{Histograms showing the distribution of the timing variable for the different methods (light blue) around their respective mean $\overline{t}$ for NO (left) and IO (right) with a Gaussian fit on top (dark blue) except for the BH collapse method which is best described by a half Breit-Wigner fit (dark blue). The shaded area shows the $1\sigma$ standard deviation around the mean timing for each method. Except for the BH collapse method, the timing variable of all other methods are well approximated by a Gaussian distribution. The different behavior for the BH method is due to the hard cutoff at the time of collapse. As an example, we show the timing distribution for the ls180s12.0 model and the BH ls220s40s7b2c model for the BH Collapse both in Hyper-Kamiokande. The normal ordering scenario is shown in the left half of the figure while the right half displays the inverted ordering scenario. For the BH Collapse method the timing was separated into $1\,\mu$s bins while all other methods were separated into bins of $200\,\mu$s. Note that the fits are only displayed for the purpose of comparison. They are not used in our determination of the timing resolution.}\label{timing_distribution}
\end{figure*}
\begin{figure*}
\includegraphics[width=\textwidth]{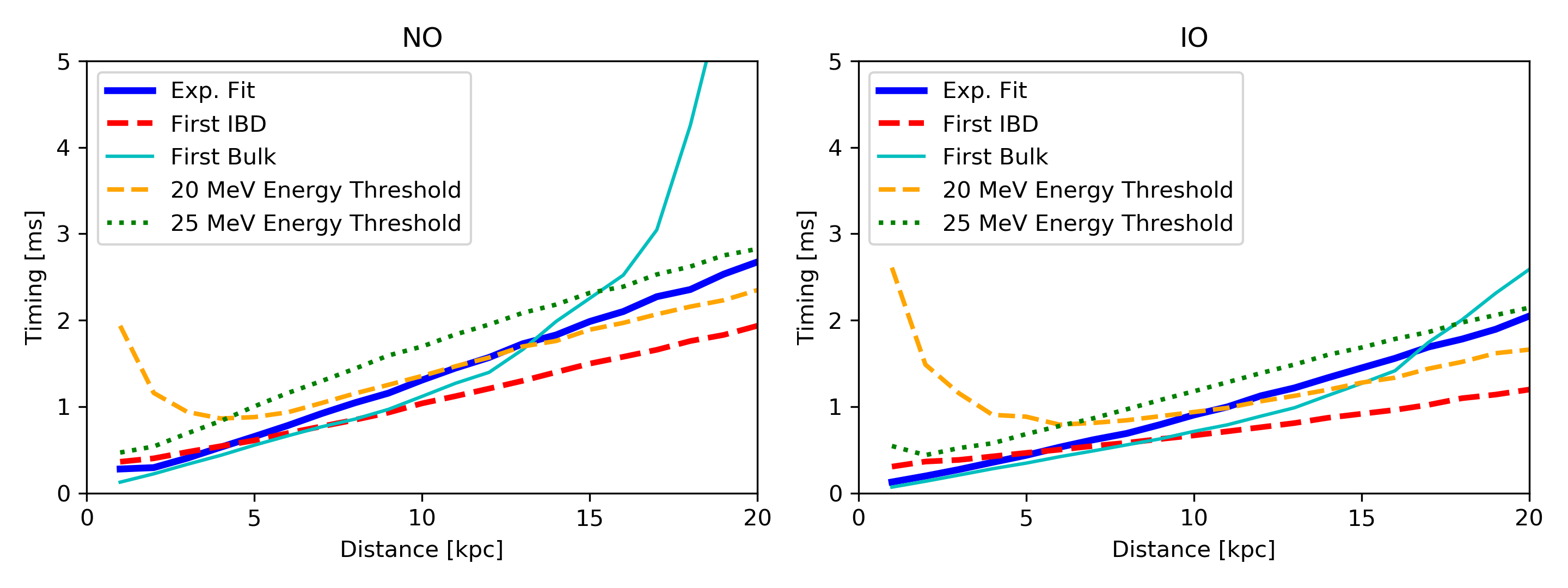}
\caption{Distance dependency of the timing resolution for four of the five methods studied to time the onset of the signal. The plot shows the different timing values for the ls180s12.0 model between $1$ and $20\,$kpc. One can see that for close supernovae below $5-7\,$kpc, the $20\,$MeV energy threshold is no longer sufficient and a larger threshold of$25\,$MeV is necessary. The opposite is true for the First Bulk method which is limited by low statistics for large distances.}\label{distance_plot}
\end{figure*}
\subsection{Timing Results}
The averaged 1$\sigma$ timing uncertainty results are shown in TABLE~\ref{avrg_results}
while the full set of results including the absolute timing off-set from core bounce for each SN simulation and each timing method are shown in Appendix~\ref{App. A} in FIG.~\ref{fig:8} and~\ref{fig:9} as well as in TABLES~\ref{table:1},~\ref{table:2},~\ref{table:3} and~\ref{table:4}.
The timing distributions for each method obtained from the ls180s12.0 and BH ls220s40s7b2c models in HK are shown in FIG.~\ref{timing_distribution}. Except for the BH method which is strongly influenced by the hard cutoff, they follow a Gaussian-like distribution. In addition, the Gauss Fit has significant non-Gaussian tails. These would be even more prominent if some of the fits had not been rejected. To summarize, the standard deviation gives a quite good description of the distributions, so little information is lost when reporting our results in the tables.\\
In FIG.~\ref{distance_plot} we show how the timing resolution based on the ls180s12.0 model develops over distance in HK. For this we simulated the signal and timing for distances between $1$ and $20\,$kpc in steps of $1\,$kpc. The Gauss Fit of the initial $\nu_e$ burst was not included in the plot due to its bad timing and to keep the plot as clean as possible. \\
As it was pointed out in Section~\ref{sec:background}, the background event rates are very low for all detectors, and to highlight this, the effect on the different timing methods can be considered: For the Exponential Fit, the fit range of $100\,$ms implies that up to $1\%$ of the MC realisations will be affected. However, a single background event would not change the fit appreciably, and the results are therefore robust. The Gauss Fit has very optimistic assumptions, so adding the assumption of no background will not change the conclusions that can be drawn. For the methods First Bulk, Energy Threshold and First IBD, a single background event could in principle change the results of one MC realisation significantly, but the background event would have to be within the first 10ms. In HK where the background rate is highest, this will occur for $0.1\%$ of the cases, and hence cannot affect the overall averaged results.
\\
The timing of First IBD and Exponential Fit increase almost linearly with distance demonstrating a $\sqrt{N}$ dependence on the number of events in the detector. The First Bulk method looses linearity at $\sim 10$kpc which is the distance it was optimised for. Using a larger integration time would make it more competitive also for larger distances. The Energy Threshold method has the opposite problem that too many neutrinos exceed the threshold for short distances. When raising $E_{\rm threshold}$ to $25$MeV, this problem is mitigated at the cost of an overall worse timing.\\
Over all, the First IBD method is performing the best although Exponential Fit and First Bulk are slightly better at small distances.\\
One important aspect to further inspect is how the different neutrino energies will affect these results if the neutrinos have non-negligible masses. 
The main effect of massive neutrinos is to shift the total signal a few ms away from the core bounce. In addition to that, there could be some influence on the shape of the signal due to ToF differences resulting from different energies of the detected neutrinos. \\
Looking at FIG.~\ref{Energyevolution}, we see that with time evolving, the mean neutrino energy increases such that for massive neutrinos, we would expect the total signal to be compressed slightly. In the case of a black hole formation, however, the hard cut-off at the end of the signal could (depending on the mass scale) become a smooth transition.
\\To inspect the effect of non-zero neutrino masses, we take the current upper limit on the effective (anti-) electron-neutrino mass from tritium decay experiments~\citep{Aseev:2011dq}. (Very recently, KATRIN has released first results putting a limit of $1.1$eV at 90\% confidence level on the electron-neutrino mass \citep{Aker:2019uuj}.) Thus,
\begin{align}
m_\nu^\text{max}=2\,\text{eV}
\end{align}
and simulate the signal shift resulting from differences in the time of flights (ToF) for different neutrino masses down to the theoretical lower limit for the heaviest mass eigenstate in the three flavor mixing scheme
\begin{align}
m_{\nu}^{\text{min}}=\sqrt{\Delta m_{32}^2}\approx0.05\,\text{eV}\;\;\;.
\end{align}
This is done for HK and the two models ls180s12.0 and BH ls220s40s7b2c and 1000 Monte Carlo realizations each. The results for some of these MC simulations are shown in Appendix~\ref{App. B}. As expected, the timing variables for methods at core bounce are only affected by the inclusion of neutrino masses through a shift in the mean arrival times. The BH formation cut-off, however, is significantly influenced by the energy dependent ToF in such a way that the abrupt hard cut-off in the detection rate becomes a rather smooth transition depending on the absolute mass scale. This effect is well known and already discussed in e.g.~\citep{Beacom:2000qy}.\\
Taking the most stringent cosmological limits on the sum of all neutrino masses~\citep{Ade:2015xua}
\begin{align}
\sum_i m_i < 0.17\,\text{eV}
\end{align}
into account, constrains the absolute neutrino masses to be below $0.1\,$eV. 
For this mass range, the BH timing resolution would grow at most by a factor of
$\sim 2$. All in all this supports the conclusion that the timing accuracy is not reduced significantly when considering massive neutrinos.
\section{Applications}\label{sec5}
\subsection{Triangulation}
First, we apply our results to estimate the angular resolution that could be achieved via triangulation. Locating the SN is not only relevant to allow for early astronomical observations, but it is especially important in the case of a failed SN where there is no strong optical signal. In general, by measuring the arrival time of the neutrino signal, two detectors separated by a distance $D$ can determine the position of the SN via the measured time difference $\Delta t$ to be on a cone along their axis with an opening angle $\theta$. We can easily calculate $\theta$ using the law of cosines as
\begin{align}
\cos{\theta}=\frac{\Delta t}{D}\;.
\end{align}
Consequently, the uncertainty in the angular resolution is
\begin{align}
\delta \cos{\theta}=\frac{\delta(\Delta t)}{D}\label{cosresolution} \;.
\end{align}
To exemplify which angular resolution the above timing results can achieve, we calculate it for the combination of IC Gen2 and HK in the non-BH case as well as HK and JUNO in the case of BH formation.
Applying the above Eq.~\eqref{cosresolution}, we find
\begin{align}
\delta(\cos\theta)_{\text{IC,HK}}=0.03\quad,\\
\delta(\cos\theta)_{\text{HK,JUNO,BH}}=0.01\;\;\;.
\end{align}
While the latter is limited by the relative proximity of both detectors and JUNOs relatively small size compared to HK, triangulating a SN in reality will utilize up to 4 different detectors. Thereby other promising candidates such as NO$\nu$A~\cite{Vasel:2017egd} or DUNE~\cite{Acciarri:2015uup} (located in the United States) which both will reach similar event rates as JUNO~\citep{Brdar:2018zds} come into play. The combination of the first HK tank with a possible second tank in Korea $\sim 800\,$km away would also reach resolutions similar to the HK-JUNO combination in the BH case despite the very short distance between the detectors.
\\
The actual angular resolution $\delta\theta$ will depend on the real angle $\theta$. For large and moderate angles up to $\theta\sim90^\circ$, the angular resolution is given by
\begin{align}
\delta\theta = \frac{\delta(\cos{\theta})}{\sin{\theta}} \;,
\end{align}
while for small angles around $\theta\sim0^\circ$, it is given by~\citep{Beacom:1998fj}
\begin{align}
\delta\theta=\sqrt{2\delta(\cos\theta)}\;\;\;.
\end{align}
Taking the above results on the resolution of $\cos\theta$, we can constrain the angular resolution for these examples to
\begin{align}
1.8^\circ\lesssim\delta\theta_{\text{IC,HK}}\lesssim 14.5^\circ\quad,\\
0.6^\circ\lesssim\delta\theta_{\text{HK,JUNO,BH}}\lesssim 8.4^\circ \quad.
\end{align}
In comparison, the angular resolution achieved by SK via neutrino-electron elastic scattering is $3^\circ-4^\circ$~\citep{Abe:2016waf}, while, based on the same calculations, HK's angular resolution is estimated to be $1^\circ-1.2^\circ$~\citep{Abe:2018uyc}.
\subsection{Neutrino Mass Determination}
Precise timing of the SN neutrino signal also offers a possibility to constrain neutrino masses. A conceptually easy way to constrain or even determine the masses of neutrinos is to use the above mentioned mass induced ToF difference in comparison to the ToF of the SN gravitational wave signal propagating at the speed of light. \\
In general, for a SN at distance $D$ and two signals with masses $m_i$ and $m_j$ both at energy $E$, the ToF difference is given by
\begin{align}
\Delta t_{ij}\approx5.1\,\text{ms}\left(\frac{D}{10\,\text{kpc}}\right)\left(\frac{\Delta m^2_{ij}}{1\,\text{eV}^2}\right)\left(\frac{E}{10\,\text{MeV}}\right)^{-2}\;\;\;.\label{tof}
\end{align}
Precise timing of the neutrino signal therefore allows to distinguish even small ToF differences and hence allows for precise constraints on the upper mass limit. With the largest mass squared difference between the neutrino mass eigenstates being at the order of $\Delta m^2\sim 2.5\cdot10^{-3}\,\text{eV}^2$, ToF differences between the different mass eigenstates will be at the order of $\sim 3\,\text{$\mu$s}$ for a neutrino energy of $20\,$MeV. We can therefore safely ignore them since none of the above techniques will reach such resolutions. \\
After LIGO's historical detection of GW150914~\citep{Abbott:2016blz}, gravitational wave astronomy has become reality, and Galactic CCSNe are promising candidates for such a measurement. To compare the neutrino signal with the gravitational wave signal, we need correlated structures in both. To first order, gravitational waves are produced by the second time derivative of the energy density quadrupole moment tensor.
Although it is null for spherical symmetric objects, SN simulations show that the flattening of the collapsing core due to its own rotation can induce a non-vanishing quadrupole moment high enough to produce a detectable gravitational wave signal (see e.g.~\citep{Ott:2012kr, PhysRevLett.106.161103, Dimmelmeier:2008iq}).\\
There are generally two characteristic signals of a short timescale that one can expect to see in the gravitational wave signal of a rotating SN. The first is the core bounce and the second is the collapse to a black hole. Luckily, the neutrino signal also shows characteristic structures at both these times namely the onset of the signal rise and the cut-off at BH formation time.
\\
To quantify how the above methods for finding and timing characteristic structures in the neutrino signal can be used to constrain neutrino masses, we assume that the model dependent mean timing value for each method is known. In this case, only the methods uncertainty contribute to the overall timing uncertainty. We also assume that the gravitational wave signal will be timed with a high precision such that the neutrino signal is the limiting factor.\\
To determine the constrainable masses for each method, we simulated the time shift induced by different non-zero neutrino masses from $0.05\,$eV up to $2\,$eV in steps of $0.1\,$eV in the range of $0.1-2.0\,$eV for all models with each 1000 realizations and determined the lowest mass that could be distinguished from zero at $90\%$ confidence level in at least $90\%$ of the MC realizations. The averaged results for HK are shown in TABLE~\ref{masstable} and selected simulations are in Appendix~\ref{App. B}.
We can compare these mass limits to possible limits resulting from a likelihood analysis~\citep{Pagliaroli:2010ik}. For SK, this analysis allows to constrain masses down to $m\sim0.8\,$eV, resulting in a possible limit of $m\sim0.45\,$eV for HK taking a scaling factor of $m^2\propto\frac{1}{\sqrt{N}}$ with $N$ being the number of detected neutrinos~\citep{Pagliaroli:2010ik}.
This comparison shows that timing single characteristic structures and their delay only gives reasonable sub-eV limits in the case of a failed SN where the timing is very precise. However, using the time delay of the exponential fit also allows IC to constrain the mass from SN neutrinos. The possible limits for IC will be comparable to HK's IBD limits.
\begin{table}[t]
\begin{tabularx}{\columnwidth}{l c Y }
\hline\hline
Method & Ordering &  {HK} \\
\hline\\
Exponential Fit [eV] & $\begin{matrix} \text{NO}\\ \text{IO} \end{matrix}$ 
& $\begin{matrix}1.7 \\ 1.4\end{matrix}$ 
\\  \\ 
First IBD [eV] & $\begin{matrix} \text{NO}\\ \text{IO} \end{matrix}$ 
& $\begin{matrix}1.4 \\ 1.0\end{matrix}$ 
\\  \\ 
First Bulk [eV] & $\begin{matrix} \text{NO}\\ \text{IO} \end{matrix}$
& $\begin{matrix}1.4 \\ 1.1\end{matrix}$ 
\\  \\ 
20 MeV Energy Threshold [eV] & $\begin{matrix} \text{NO}\\ \text{IO} \end{matrix}$
& $\begin{matrix} >2.0 \\ 1.8\end{matrix}$ 
\\  \\ 
Black Hole Collapse [eV] & $\begin{matrix} \text{NO}\\ \text{IO} \end{matrix}$
& $\begin{matrix}0.29 \\ 0.33\end{matrix}$ \\
\hline\hline
\end{tabularx}
\caption{Average neutrino masses that could be distinguished from zero at $90\%$ confidence level in at least $90\%$ of the MC realizations for each method. 
}\label{masstable}
\end{table}
\section{Conclusion}\label{sec6}
We investigated six possible methods for timing the neutrino signal of a Galactic supernova for three (five for Exponential Fit) existing and future detectors. Our results show that HK will be comparable to today's IceCube detector both being able to achieve $\sim1\,$ms precision, while in the case of a failed SN, even the smaller SK and JUNO detectors can reach sub-ms precision.
Additionally, we found that the very intuitive idea of timing the characteristic initial $\nu_e$-burst shortly after core bounce fails in most of the scenarios. The only candidate that our analysis finds could potentially see the $\nu_e$-burst is Hyper-Kamiokande. However, if the $\nu_e$-burst is detected by the future HK experiment, it would be a hint towards an inverted mass hierarchy.
Another interesting candidate for observing the $\nu_e$-burst is DUNE~\citep{Ankowski:2016lab} which should be included in future studies, once its systematics have been studied in more details.
The detectability and timing of the initial burst was also investigated by~\citep{Wallace:2015xma} using a different fitfunction. The results are comparable.\\ In the exciting case that the next Galactic supernova will fail and result in the protoneutron star collapsing to a black hole during accretion or early cooling phase time, Hyper-Kamiokande might be able to actually observe how the formation process proceeds in the neutrino signal. This will depend on the actual distance.
\\
Three methods (Exponential Fit, First Bulk, and Gauss Fit) use a fit over several neutrino events. While the latter does not work in most cases, the Exponential Fit method results in stable timings, and due to the fact that it is using many neutrino events, it is not affected by background events. The same holds for the First Bulk method. The other methods (First IBD, Energy Threshold and BH Collapse) all use the statistical fluctuations in the timing of single neutrino events making them more background sensitive. However, compared to the event rate during a SN, the background in the relevant energy range is negligible~\citep{Djurcic:2015vqa, Abe:2016waf, Abe:2018uyc}. Especially the First IBD method, due to its characteristic signature of a positron followed by neutron capture, is rather insensitive to backgrounds. One should note that for closer supernovae a slightly higher energy threshold of $25\,$MeV results in a better timing than the $20\,$MeV threshold.
\\
Comparing the different detectors, the future IceCube Gen2 update will deliver the most precise timing resolution in the non-BH case while HK, SK and JUNO allow very precise timings in the case of a BH formation. Here IceCube is again limited by the fact that it will detect a SN by noise excess rather than single events. However, with the addition of the HitSpooling system, IC will be capable of resolving the BH collapse with a resolution similar to that of HK and even provide internal triangulation of the location~\citep{HeeremanvonZuydtwyck:2015mbs, Kopke:2017req}. The improved data binning should, however, not influence the timing of the onset of the burst significantly since this is limited by the still existing noise rate rather than the data binning.\\
In the last section, we studied the impact of the timing results on two possible applications, the first being the location of the SN via triangulation. Taking the example of HK+IC, we found that, for a SN that is approximately perpendicular to the connecting axis between the two considered detectors, the angular resolution is comparable to the method of locating the SN via neutrino-electron elastic scattering. In the case of a failed SN, the HK+JUNO combination can potentially reach sub-degree resolution. Similar results can be obtained by combining the Japanese HK tank with a second Korean tank.\\
At last we studied the possibility to constrain neutrino masses via ToF differences in comparison to gravitational waves. In the non-BH case, we found that by timing the onset of the signal, HK can limit neutrino masses to $\sim1\,$eV. This improves to $\sim0.3\,$eV in the BH forming case considering a SN at $10\,$kpc. The latter result is comparable to the goal of the KATRIN experiment~\cite{Osipowicz:2001sq}.
\begin{acknowledgements}
We would like to thank Hans-Thomas Janka and Tobias Melson for giving us access to the Garching CCSN archive. R.S.L.H. was partly funded by the Alexander von Humboldt Foundation.
\end{acknowledgements}
\onecolumngrid
\clearpage
\appendix
\section{Summarized Simulations}\label{App. A}
\begin{figure}[hb]
    \includegraphics[width=0.9\textwidth]{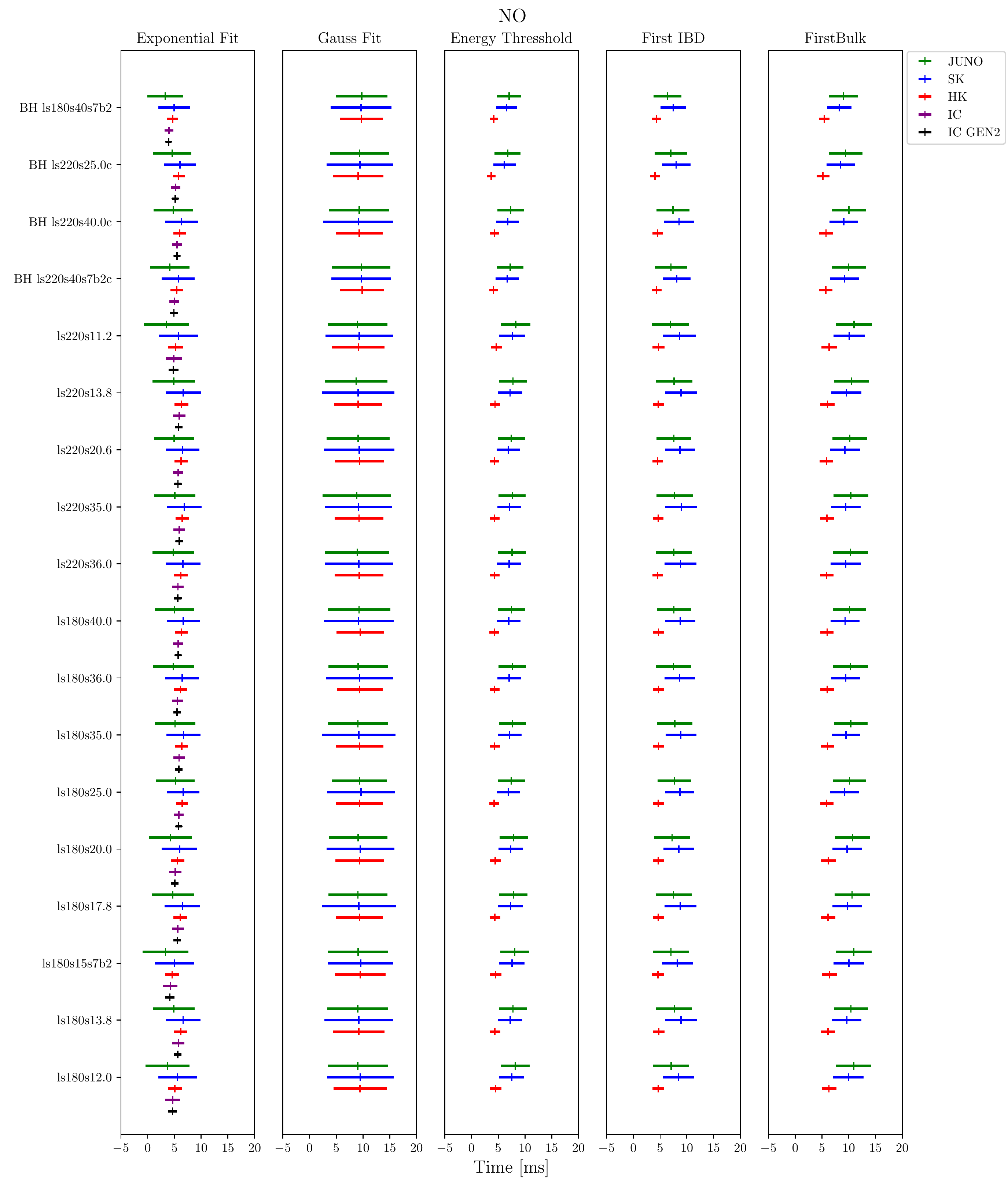}\\
    \caption{Visualization of the timing performance, i.e., average and standard deviation for each detector, method (except BH) and supernova model used assuming normal mass ordering. The plot shows the average timing compared to the core bounce at $t=0$ as well as the $1\sigma$ deviation band.}
    \label{fig:8}
\end{figure}

\begin{figure}[hb]
    \includegraphics[width=0.9\textwidth]{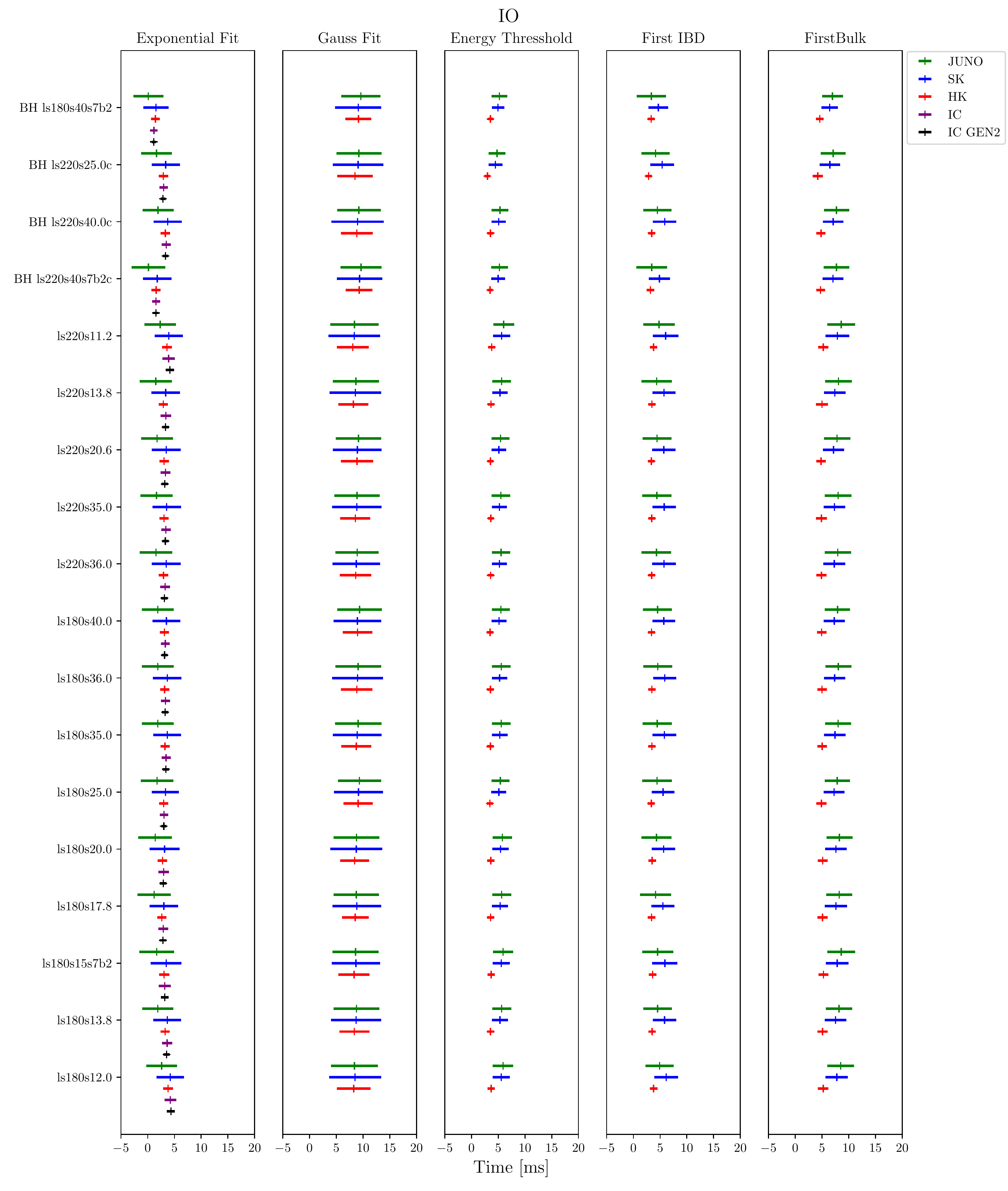}\\
    \caption{Visualization of the timing performance, i.e., average and standard deviation for each detector, method (except BH) and supernova model used assuming inverted mass ordering. The plot shows the average timing compared to the core bounce at $t=0$ as well as the $1\sigma$ deviation band.}
    \label{fig:9}
\end{figure}

Here we show the full set of results including the absolute timing off-set from core bounce for each SN simulation and each timing method are shown in FIG.~\ref{fig:8} and~\ref{fig:9} as well as in TABLEs~\ref{table:1},~\ref{table:2},~\ref{table:3} and~\ref{table:4}.
\newpage
\newpage
\begin{table}[]
\begin{tabularx}{\textwidth}{l l Y Y Y Y Y Y Y}
\hline\hline\\
 \multicolumn{9}{c}{SK - All Simulations}\\\\
Simulation &  & $\begin{matrix}
\text{Exponential}\\
\text{Fit}\\
\text{[ms]}
\end{matrix}$ & $\begin{matrix}
\text{}\\
\text{Gauss Fit}\\
\text{[ms]}
\end{matrix}$ & $\begin{matrix}
\text{failed}\\
\text{Gauss Fit}\\
\text{[\%]}
\end{matrix}$ & $\begin{matrix}
\text{Energy}\\
\text{Threshold}\\
\text{($20\,$MeV) [ms]}
\end{matrix}$ & $\begin{matrix}
\text{}\\
\text{First IBD}\\
\text{[ms]}
\end{matrix}$ & $\begin{matrix}
\text{}\\
\text{First Bulk}\\
\text{[ms]}
\end{matrix}$ & $\begin{matrix}
\text{BH}\\
\text{Collapse}\\
\text{[ms]}
\end{matrix}$ \\
\hline\\
ls180s12.0 & $\begin{matrix} \text{NO}\\ \text{IO} \end{matrix}$ & $\begin{matrix} 5.6\pm 3.6 \\ 4.2\pm 2.6\end{matrix}$ & $ \begin{matrix} 9.5\pm 6.2\\ 8.5\pm 4.9\end{matrix}$ & $ \begin{matrix} 62 \\ 48\end{matrix}$ & $\begin{matrix} 9.9\pm 2.9 \\ 7.8\pm 2.1\end{matrix}$ & $\begin{matrix} 7.5\pm 2.4 \\ 5.6\pm 1.6\end{matrix}$ & $ \begin{matrix} 8.4\pm 2.9 \\ 5.6\pm 2.2\end{matrix}$ & $ \begin{matrix} -  \\ - \end{matrix}$ \\  \\ 
ls180s13.8 & $\begin{matrix} \text{NO}\\ \text{IO} \end{matrix}$ & $\begin{matrix} 6.6\pm 3.3 \\ 3.7\pm 2.6\end{matrix}$ & $ \begin{matrix} 9.2\pm 6.4\\ 8.7\pm 4.7\end{matrix}$ & $ \begin{matrix} 60 \\ 48\end{matrix}$ & $\begin{matrix} 9.6\pm 2.8 \\ 7.5\pm 2.0\end{matrix}$ & $\begin{matrix} 7.2\pm 2.3 \\ 5.3\pm 1.5\end{matrix}$ & $ \begin{matrix} 8.9\pm 2.9 \\ 5.3\pm 2.2\end{matrix}$ & $ \begin{matrix} -  \\ - \end{matrix}$ \\  \\ 
ls180s15s7b2 & $\begin{matrix} \text{NO}\\ \text{IO} \end{matrix}$ & $\begin{matrix} 5.0\pm 3.6 \\ 3.5\pm 2.9\end{matrix}$ & $ \begin{matrix} 9.6\pm 6.1\\ 8.6\pm 4.5\end{matrix}$ & $ \begin{matrix} 63 \\ 49\end{matrix}$ & $\begin{matrix} 10.0\pm 2.9 \\ 7.8\pm 2.1\end{matrix}$ & $\begin{matrix} 7.6\pm 2.4 \\ 5.6\pm 1.6\end{matrix}$ & $ \begin{matrix} 8.2\pm 2.9 \\ 5.6\pm 2.4\end{matrix}$ & $ \begin{matrix} -  \\ - \end{matrix}$ \\  \\ 
ls180s17.8 & $\begin{matrix} \text{NO}\\ \text{IO} \end{matrix}$ & $\begin{matrix} 6.5\pm 3.3 \\ 3.0\pm 2.7\end{matrix}$ & $ \begin{matrix} 9.2\pm 6.9\\ 8.8\pm 4.6\end{matrix}$ & $ \begin{matrix} 60 \\ 50\end{matrix}$ & $\begin{matrix} 9.7\pm 2.8 \\ 7.6\pm 2.1\end{matrix}$ & $\begin{matrix} 7.3\pm 2.3 \\ 5.3\pm 1.5\end{matrix}$ & $ \begin{matrix} 8.8\pm 3.0 \\ 5.3\pm 2.2\end{matrix}$ & $ \begin{matrix} -  \\ - \end{matrix}$ \\  \\ 
ls180s20.0 & $\begin{matrix} \text{NO}\\ \text{IO} \end{matrix}$ & $\begin{matrix} 6.0\pm 3.3 \\ 3.2\pm 2.8\end{matrix}$ & $ \begin{matrix} 9.5\pm 6.3\\ 8.7\pm 4.9\end{matrix}$ & $ \begin{matrix} 61 \\ 49\end{matrix}$ & $\begin{matrix} 9.7\pm 2.8 \\ 7.6\pm 2.0\end{matrix}$ & $\begin{matrix} 7.4\pm 2.3 \\ 5.4\pm 1.5\end{matrix}$ & $ \begin{matrix} 8.5\pm 2.9 \\ 5.4\pm 2.2\end{matrix}$ & $ \begin{matrix} -  \\ - \end{matrix}$ \\  \\ 
ls180s25.0 & $\begin{matrix} \text{NO}\\ \text{IO} \end{matrix}$ & $\begin{matrix} 6.7\pm 3.0 \\ 3.3\pm 2.5\end{matrix}$ & $ \begin{matrix} 9.6\pm 6.3\\ 9.1\pm 4.6\end{matrix}$ & $ \begin{matrix} 58 \\ 49\end{matrix}$ & $\begin{matrix} 9.2\pm 2.7 \\ 7.2\pm 2.0\end{matrix}$ & $\begin{matrix} 6.9\pm 2.2 \\ 5.1\pm 1.4\end{matrix}$ & $ \begin{matrix} 8.7\pm 2.7 \\ 5.1\pm 2.1\end{matrix}$ & $ \begin{matrix} -  \\ - \end{matrix}$ \\  \\ 
ls180s35.0 & $\begin{matrix} \text{NO}\\ \text{IO} \end{matrix}$ & $\begin{matrix} 6.7\pm 3.2 \\ 3.7\pm 2.6\end{matrix}$ & $ \begin{matrix} 9.2\pm 6.9\\ 8.9\pm 4.6\end{matrix}$ & $ \begin{matrix} 59 \\ 49\end{matrix}$ & $\begin{matrix} 9.5\pm 2.7 \\ 7.4\pm 2.0\end{matrix}$ & $\begin{matrix} 7.1\pm 2.2 \\ 5.3\pm 1.5\end{matrix}$ & $ \begin{matrix} 8.9\pm 2.9 \\ 5.3\pm 2.2\end{matrix}$ & $ \begin{matrix} -  \\ - \end{matrix}$ \\  \\ 
ls180s36.0 & $\begin{matrix} \text{NO}\\ \text{IO} \end{matrix}$ & $\begin{matrix} 6.4\pm 3.2 \\ 3.7\pm 2.7\end{matrix}$ & $ \begin{matrix} 9.4\pm 6.3\\ 9.0\pm 4.7\end{matrix}$ & $ \begin{matrix} 59 \\ 48\end{matrix}$ & $\begin{matrix} 9.5\pm 2.7 \\ 7.4\pm 2.0\end{matrix}$ & $\begin{matrix} 7.0\pm 2.2 \\ 5.3\pm 1.5\end{matrix}$ & $ \begin{matrix} 8.7\pm 2.8 \\ 5.3\pm 2.1\end{matrix}$ & $ \begin{matrix} -  \\ - \end{matrix}$ \\  \\ 
ls180s40.0 & $\begin{matrix} \text{NO}\\ \text{IO} \end{matrix}$ & $\begin{matrix} 6.7\pm 3.1 \\ 3.5\pm 2.6\end{matrix}$ & $ \begin{matrix} 9.2\pm 6.5\\ 8.9\pm 4.4\end{matrix}$ & $ \begin{matrix} 58 \\ 50\end{matrix}$ & $\begin{matrix} 9.3\pm 2.7 \\ 7.3\pm 2.0\end{matrix}$ & $\begin{matrix} 7.0\pm 2.2 \\ 5.2\pm 1.4\end{matrix}$ & $ \begin{matrix} 8.8\pm 2.8 \\ 5.2\pm 2.1\end{matrix}$ & $ \begin{matrix} -  \\ - \end{matrix}$ \\  \\ 
ls220s11.2 & $\begin{matrix} \text{NO}\\ \text{IO} \end{matrix}$ & $\begin{matrix} 5.7\pm 3.6 \\ 4.0\pm 2.6\end{matrix}$ & $ \begin{matrix} 9.3\pm 6.3\\ 8.3\pm 4.8\end{matrix}$ & $ \begin{matrix} 63 \\ 47\end{matrix}$ & $\begin{matrix} 10.1\pm 2.9 \\ 7.9\pm 2.2\end{matrix}$ & $\begin{matrix} 7.6\pm 2.4 \\ 5.6\pm 1.6\end{matrix}$ & $ \begin{matrix} 8.6\pm 3.0 \\ 5.6\pm 2.4\end{matrix}$ & $ \begin{matrix} -  \\ - \end{matrix}$ \\  \\ 
ls220s13.8 & $\begin{matrix} \text{NO}\\ \text{IO} \end{matrix}$ & $\begin{matrix} 6.7\pm 3.3 \\ 3.4\pm 2.7\end{matrix}$ & $ \begin{matrix} 9.1\pm 6.8\\ 8.6\pm 4.8\end{matrix}$ & $ \begin{matrix} 60 \\ 49\end{matrix}$ & $\begin{matrix} 9.6\pm 2.8 \\ 7.4\pm 2.0\end{matrix}$ & $\begin{matrix} 7.2\pm 2.3 \\ 5.3\pm 1.4\end{matrix}$ & $ \begin{matrix} 8.9\pm 3.0 \\ 5.3\pm 2.2\end{matrix}$ & $ \begin{matrix} -  \\ - \end{matrix}$ \\  \\ 
ls220s20.6 & $\begin{matrix} \text{NO}\\ \text{IO} \end{matrix}$ & $\begin{matrix} 6.5\pm 3.1 \\ 3.4\pm 2.7\end{matrix}$ & $ \begin{matrix} 9.3\pm 6.6\\ 8.9\pm 4.6\end{matrix}$ & $ \begin{matrix} 56 \\ 50\end{matrix}$ & $\begin{matrix} 9.3\pm 2.8 \\ 7.2\pm 2.0\end{matrix}$ & $\begin{matrix} 6.9\pm 2.2 \\ 5.1\pm 1.4\end{matrix}$ & $ \begin{matrix} 8.7\pm 2.8 \\ 5.1\pm 2.2\end{matrix}$ & $ \begin{matrix} -  \\ - \end{matrix}$ \\  \\ 
ls220s35.0 & $\begin{matrix} \text{NO}\\ \text{IO} \end{matrix}$ & $\begin{matrix} 6.8\pm 3.3 \\ 3.5\pm 2.7\end{matrix}$ & $ \begin{matrix} 9.2\pm 6.3\\ 8.8\pm 4.6\end{matrix}$ & $ \begin{matrix} 58 \\ 48\end{matrix}$ & $\begin{matrix} 9.5\pm 2.8 \\ 7.3\pm 2.0\end{matrix}$ & $\begin{matrix} 7.1\pm 2.2 \\ 5.2\pm 1.4\end{matrix}$ & $ \begin{matrix} 9.0\pm 3.0 \\ 5.2\pm 2.2\end{matrix}$ & $ \begin{matrix} -  \\ - \end{matrix}$ \\  \\ 
ls220s36.0 & $\begin{matrix} \text{NO}\\ \text{IO} \end{matrix}$ & $\begin{matrix} 6.6\pm 3.2 \\ 3.5\pm 2.7\end{matrix}$ & $ \begin{matrix} 9.2\pm 6.4\\ 8.7\pm 4.5\end{matrix}$ & $ \begin{matrix} 60 \\ 49\end{matrix}$ & $\begin{matrix} 9.4\pm 2.8 \\ 7.3\pm 2.0\end{matrix}$ & $\begin{matrix} 7.0\pm 2.3 \\ 5.2\pm 1.4\end{matrix}$ & $ \begin{matrix} 8.8\pm 3.0 \\ 5.2\pm 2.2\end{matrix}$ & $ \begin{matrix} -  \\ - \end{matrix}$ \\  \\ 
BH ls180s40s7b2 & $\begin{matrix} \text{NO}\\ \text{IO} \end{matrix}$ & $\begin{matrix} 4.9\pm 2.9 \\ 1.5\pm 2.4\end{matrix}$ & $ \begin{matrix} 9.6\pm 5.7\\ 9.1\pm 4.3\end{matrix}$ & $ \begin{matrix} 58 \\ 49\end{matrix}$ & $\begin{matrix} 8.2\pm 2.3 \\ 6.4\pm 1.5\end{matrix}$ & $\begin{matrix} 6.6\pm 1.9 \\ 5.0\pm 1.2\end{matrix}$ & $ \begin{matrix} 7.5\pm 2.4 \\ 5.0\pm 1.8\end{matrix}$ & $ \begin{matrix} 434.93\pm0.05 \\ 434.93\pm0.05\end{matrix}$ \\  \\ 
BH ls220s25.0c & $\begin{matrix} \text{NO}\\ \text{IO} \end{matrix}$ & $\begin{matrix} 6.0\pm 3.0 \\ 3.4\pm 2.6\end{matrix}$ & $ \begin{matrix} 9.4\pm 6.2\\ 9.1\pm 4.7\end{matrix}$ & $ \begin{matrix} 58 \\ 50\end{matrix}$ & $\begin{matrix} 8.5\pm 2.6 \\ 6.5\pm 1.9\end{matrix}$ & $\begin{matrix} 6.2\pm 2.1 \\ 4.5\pm 1.3\end{matrix}$ & $ \begin{matrix} 8.0\pm 2.7 \\ 4.5\pm 2.2\end{matrix}$ & $ \begin{matrix} 1277.34\pm0.11 \\ 1277.3\pm0.15\end{matrix}$ \\  \\ 
BH ls220s40.0c & $\begin{matrix} \text{NO}\\ \text{IO} \end{matrix}$ & $\begin{matrix} 6.3\pm 3.1 \\ 3.8\pm 2.6\end{matrix}$ & $ \begin{matrix} 9.1\pm 6.6\\ 9.0\pm 4.9\end{matrix}$ & $ \begin{matrix} 59 \\ 49\end{matrix}$ & $\begin{matrix} 9.1\pm 2.7 \\ 7.1\pm 1.9\end{matrix}$ & $\begin{matrix} 6.8\pm 2.1 \\ 5.1\pm 1.3\end{matrix}$ & $ \begin{matrix} 8.5\pm 2.8 \\ 5.1\pm 2.2\end{matrix}$ & $ \begin{matrix} 2105.48\pm0.12 \\ 2105.41\pm0.19\end{matrix}$ \\  \\ 
BH ls220s40s7b2c & $\begin{matrix} \text{NO}\\ \text{IO} \end{matrix}$ & $\begin{matrix} 5.7\pm 3.1 \\ 1.8\pm 2.7\end{matrix}$ & $ \begin{matrix} 9.7\pm 5.6\\ 9.3\pm 4.2\end{matrix}$ & $ \begin{matrix} 59 \\ 50\end{matrix}$ & $\begin{matrix} 9.2\pm 2.7 \\ 7.0\pm 2.0\end{matrix}$ & $\begin{matrix} 6.7\pm 2.2 \\ 5.0\pm 1.3\end{matrix}$ & $ \begin{matrix} 8.1\pm 2.6 \\ 5.0\pm 2.0\end{matrix}$ & $ \begin{matrix} 567.88\pm0.05 \\ 567.88\pm0.05\end{matrix}$ \\  \\

\hline\hline
\end{tabularx}
\caption{Summary of the results of the different timing methods in Super-Kamiokande for each of the 18 SN models used.}\label{table:1}
\end{table}
\begin{table}[]
\begin{tabularx}{\textwidth}{l l Y Y Y Y Y Y Y}
\hline\hline\\
 \multicolumn{9}{c}{JUNO - All Simulations}\\\\
Simulation & & $\begin{matrix}
\text{Exponential}\\
\text{Fit}\\
\text{[ms]}
\end{matrix}$ & $\begin{matrix}
\text{}\\
\text{Gauss Fit}\\
\text{[ms]}
\end{matrix}$ & $\begin{matrix}
\text{failed}\\
\text{Gauss Fit}\\
\text{[\%]}
\end{matrix}$ & $\begin{matrix}
\text{Energy}\\
\text{Threshold}\\
\text{($20\,$MeV) [ms]}
\end{matrix}$ & $\begin{matrix}
\text{}\\
\text{First IBD}\\
\text{[ms]}
\end{matrix}$ & $\begin{matrix}
\text{}\\
\text{First Bulk}\\
\text{[ms]}
\end{matrix}$ & $\begin{matrix}
\text{BH}\\
\text{Collapse}\\
\text{[ms]}
\end{matrix}$ \\
\hline\\
ls180s12.0 & $\begin{matrix} \text{NO}\\ \text{IO} \end{matrix}$ & $\begin{matrix} 3.7\pm 4.1 \\ 2.6\pm 2.9\end{matrix}$ & $ \begin{matrix} 9.0\pm 5.6\\ 8.4\pm 4.4\end{matrix}$ & $ \begin{matrix} 48 \\ 46\end{matrix}$ & $\begin{matrix} 10.9\pm 3.3 \\ 8.5\pm 2.5\end{matrix}$ & $\begin{matrix} 8.2\pm 2.7 \\ 5.9\pm 1.9\end{matrix}$ & $ \begin{matrix} 7.1\pm 3.3 \\ 5.9\pm 2.7\end{matrix}$ & $ \begin{matrix} -  \\ - \end{matrix}$ \\  \\ 
ls180s13.8 & $\begin{matrix} \text{NO}\\ \text{IO} \end{matrix}$ & $\begin{matrix} 4.9\pm 3.9 \\ 1.9\pm 2.9\end{matrix}$ & $ \begin{matrix} 9.0\pm 5.7\\ 8.8\pm 4.3\end{matrix}$ & $ \begin{matrix} 44 \\ 44\end{matrix}$ & $\begin{matrix} 10.4\pm 3.2 \\ 8.2\pm 2.5\end{matrix}$ & $\begin{matrix} 7.8\pm 2.6 \\ 5.6\pm 1.8\end{matrix}$ & $ \begin{matrix} 7.6\pm 3.4 \\ 5.6\pm 2.7\end{matrix}$ & $ \begin{matrix} -  \\ - \end{matrix}$ \\  \\ 
ls180s15s7b2 & $\begin{matrix} \text{NO}\\ \text{IO} \end{matrix}$ & $\begin{matrix} 3.3\pm 4.3 \\ 1.7\pm 3.2\end{matrix}$ & $ \begin{matrix} 9.1\pm 5.6\\ 8.6\pm 4.3\end{matrix}$ & $ \begin{matrix} 48 \\ 46\end{matrix}$ & $\begin{matrix} 10.9\pm 3.4 \\ 8.6\pm 2.6\end{matrix}$ & $\begin{matrix} 8.1\pm 2.7 \\ 5.9\pm 1.9\end{matrix}$ & $ \begin{matrix} 7.0\pm 3.3 \\ 5.9\pm 2.9\end{matrix}$ & $ \begin{matrix} -  \\ - \end{matrix}$ \\  \\ 
ls180s17.8 & $\begin{matrix} \text{NO}\\ \text{IO} \end{matrix}$ & $\begin{matrix} 4.7\pm 4.0 \\ 1.2\pm 3.1\end{matrix}$ & $ \begin{matrix} 9.0\pm 5.5\\ 8.7\pm 4.3\end{matrix}$ & $ \begin{matrix} 45 \\ 45\end{matrix}$ & $\begin{matrix} 10.6\pm 3.3 \\ 8.2\pm 2.4\end{matrix}$ & $\begin{matrix} 7.8\pm 2.7 \\ 5.6\pm 1.8\end{matrix}$ & $ \begin{matrix} 7.5\pm 3.4 \\ 5.6\pm 2.9\end{matrix}$ & $ \begin{matrix} -  \\ - \end{matrix}$ \\  \\ 
ls180s20.0 & $\begin{matrix} \text{NO}\\ \text{IO} \end{matrix}$ & $\begin{matrix} 4.3\pm 4.0 \\ 1.4\pm 3.2\end{matrix}$ & $ \begin{matrix} 9.1\pm 5.4\\ 8.8\pm 4.3\end{matrix}$ & $ \begin{matrix} 46 \\ 45\end{matrix}$ & $\begin{matrix} 10.7\pm 3.3 \\ 8.3\pm 2.4\end{matrix}$ & $\begin{matrix} 7.9\pm 2.6 \\ 5.8\pm 1.8\end{matrix}$ & $ \begin{matrix} 7.2\pm 3.3 \\ 5.8\pm 2.8\end{matrix}$ & $ \begin{matrix} -  \\ - \end{matrix}$ \\  \\ 
ls180s25.0 & $\begin{matrix} \text{NO}\\ \text{IO} \end{matrix}$ & $\begin{matrix} 5.2\pm 3.6 \\ 1.8\pm 3.0\end{matrix}$ & $ \begin{matrix} 9.4\pm 5.2\\ 9.3\pm 4.0\end{matrix}$ & $ \begin{matrix} 43 \\ 43\end{matrix}$ & $\begin{matrix} 10.1\pm 3.1 \\ 7.8\pm 2.4\end{matrix}$ & $\begin{matrix} 7.4\pm 2.5 \\ 5.4\pm 1.7\end{matrix}$ & $ \begin{matrix} 7.7\pm 3.1 \\ 5.4\pm 2.8\end{matrix}$ & $ \begin{matrix} -  \\ - \end{matrix}$ \\  \\ 
ls180s35.0 & $\begin{matrix} \text{NO}\\ \text{IO} \end{matrix}$ & $\begin{matrix} 5.1\pm 3.8 \\ 1.9\pm 3.0\end{matrix}$ & $ \begin{matrix} 9.0\pm 5.6\\ 9.1\pm 4.4\end{matrix}$ & $ \begin{matrix} 44 \\ 43\end{matrix}$ & $\begin{matrix} 10.4\pm 3.2 \\ 8.0\pm 2.5\end{matrix}$ & $\begin{matrix} 7.7\pm 2.5 \\ 5.6\pm 1.7\end{matrix}$ & $ \begin{matrix} 7.7\pm 3.3 \\ 5.6\pm 2.7\end{matrix}$ & $ \begin{matrix} -  \\ - \end{matrix}$ \\  \\ 
ls180s36.0 & $\begin{matrix} \text{NO}\\ \text{IO} \end{matrix}$ & $\begin{matrix} 4.8\pm 3.8 \\ 1.9\pm 3.0\end{matrix}$ & $ \begin{matrix} 9.1\pm 5.6\\ 9.1\pm 4.3\end{matrix}$ & $ \begin{matrix} 45 \\ 43\end{matrix}$ & $\begin{matrix} 10.4\pm 3.3 \\ 8.1\pm 2.4\end{matrix}$ & $\begin{matrix} 7.6\pm 2.6 \\ 5.6\pm 1.7\end{matrix}$ & $ \begin{matrix} 7.5\pm 3.3 \\ 5.6\pm 2.7\end{matrix}$ & $ \begin{matrix} -  \\ - \end{matrix}$ \\  \\ 
ls180s40.0 & $\begin{matrix} \text{NO}\\ \text{IO} \end{matrix}$ & $\begin{matrix} 5.0\pm 3.7 \\ 1.9\pm 3.0\end{matrix}$ & $ \begin{matrix} 9.2\pm 5.9\\ 9.4\pm 4.2\end{matrix}$ & $ \begin{matrix} 44 \\ 44\end{matrix}$ & $\begin{matrix} 10.2\pm 3.1 \\ 7.9\pm 2.4\end{matrix}$ & $\begin{matrix} 7.5\pm 2.5 \\ 5.5\pm 1.7\end{matrix}$ & $ \begin{matrix} 7.6\pm 3.2 \\ 5.5\pm 2.7\end{matrix}$ & $ \begin{matrix} -  \\ - \end{matrix}$ \\  \\ 
ls220s11.2 & $\begin{matrix} \text{NO}\\ \text{IO} \end{matrix}$ & $\begin{matrix} 3.5\pm 4.2 \\ 2.3\pm 2.9\end{matrix}$ & $ \begin{matrix} 9.0\pm 5.6\\ 8.4\pm 4.5\end{matrix}$ & $ \begin{matrix} 48 \\ 45\end{matrix}$ & $\begin{matrix} 11.0\pm 3.4 \\ 8.6\pm 2.6\end{matrix}$ & $\begin{matrix} 8.3\pm 2.8 \\ 6.0\pm 2.0\end{matrix}$ & $ \begin{matrix} 7.0\pm 3.5 \\ 6.0\pm 3.0\end{matrix}$ & $ \begin{matrix} -  \\ - \end{matrix}$ \\  \\ 
ls220s13.8 & $\begin{matrix} \text{NO}\\ \text{IO} \end{matrix}$ & $\begin{matrix} 4.9\pm 4.0 \\ 1.5\pm 3.0\end{matrix}$ & $ \begin{matrix} 8.7\pm 5.9\\ 8.7\pm 4.3\end{matrix}$ & $ \begin{matrix} 46 \\ 43\end{matrix}$ & $\begin{matrix} 10.5\pm 3.3 \\ 8.1\pm 2.5\end{matrix}$ & $\begin{matrix} 7.7\pm 2.7 \\ 5.6\pm 1.7\end{matrix}$ & $ \begin{matrix} 7.6\pm 3.4 \\ 5.6\pm 2.9\end{matrix}$ & $ \begin{matrix} -  \\ - \end{matrix}$ \\  \\ 
ls220s20.6 & $\begin{matrix} \text{NO}\\ \text{IO} \end{matrix}$ & $\begin{matrix} 5.0\pm 3.8 \\ 1.8\pm 3.0\end{matrix}$ & $ \begin{matrix} 9.1\pm 5.9\\ 9.1\pm 4.3\end{matrix}$ & $ \begin{matrix} 44 \\ 43\end{matrix}$ & $\begin{matrix} 10.2\pm 3.2 \\ 7.8\pm 2.5\end{matrix}$ & $\begin{matrix} 7.5\pm 2.5 \\ 5.4\pm 1.7\end{matrix}$ & $ \begin{matrix} 7.6\pm 3.2 \\ 5.4\pm 2.7\end{matrix}$ & $ \begin{matrix} -  \\ - \end{matrix}$ \\  \\ 
ls220s35.0 & $\begin{matrix} \text{NO}\\ \text{IO} \end{matrix}$ & $\begin{matrix} 5.1\pm 3.9 \\ 1.6\pm 3.0\end{matrix}$ & $ \begin{matrix} 8.8\pm 6.4\\ 8.9\pm 4.3\end{matrix}$ & $ \begin{matrix} 44 \\ 43\end{matrix}$ & $\begin{matrix} 10.4\pm 3.3 \\ 8.0\pm 2.5\end{matrix}$ & $\begin{matrix} 7.6\pm 2.5 \\ 5.5\pm 1.7\end{matrix}$ & $ \begin{matrix} 7.7\pm 3.4 \\ 5.5\pm 2.8\end{matrix}$ & $ \begin{matrix} -  \\ - \end{matrix}$ \\  \\ 
ls220s36.0 & $\begin{matrix} \text{NO}\\ \text{IO} \end{matrix}$ & $\begin{matrix} 4.8\pm 3.9 \\ 1.6\pm 3.0\end{matrix}$ & $ \begin{matrix} 8.9\pm 6.0\\ 8.9\pm 4.0\end{matrix}$ & $ \begin{matrix} 45 \\ 43\end{matrix}$ & $\begin{matrix} 10.3\pm 3.2 \\ 8.0\pm 2.5\end{matrix}$ & $\begin{matrix} 7.6\pm 2.6 \\ 5.5\pm 1.7\end{matrix}$ & $ \begin{matrix} 7.5\pm 3.4 \\ 5.5\pm 2.8\end{matrix}$ & $ \begin{matrix} -  \\ - \end{matrix}$ \\  \\ 
BH ls180s40s7b2 & $\begin{matrix} \text{NO}\\ \text{IO} \end{matrix}$ & $\begin{matrix} 3.3\pm 3.3 \\ 0.1\pm 2.8\end{matrix}$ & $ \begin{matrix} 9.8\pm 4.8\\ 9.6\pm 3.7\end{matrix}$ & $ \begin{matrix} 45 \\ 44\end{matrix}$ & $\begin{matrix} 9.0\pm 2.7 \\ 7.0\pm 2.0\end{matrix}$ & $\begin{matrix} 7.0\pm 2.3 \\ 5.2\pm 1.5\end{matrix}$ & $ \begin{matrix} 6.4\pm 2.6 \\ 5.2\pm 2.8\end{matrix}$ & $ \begin{matrix} 434.92\pm0.06 \\ 434.92\pm0.06\end{matrix}$ \\  \\ 
BH ls220s25.0c & $\begin{matrix} \text{NO}\\ \text{IO} \end{matrix}$ & $\begin{matrix} 4.6\pm 3.6 \\ 1.7\pm 2.9\end{matrix}$ & $ \begin{matrix} 9.4\pm 5.5\\ 9.2\pm 4.2\end{matrix}$ & $ \begin{matrix} 44 \\ 44\end{matrix}$ & $\begin{matrix} 9.4\pm 3.2 \\ 7.1\pm 2.3\end{matrix}$ & $\begin{matrix} 6.8\pm 2.4 \\ 4.8\pm 1.6\end{matrix}$ & $ \begin{matrix} 7.0\pm 3.0 \\ 4.8\pm 2.6\end{matrix}$ & $ \begin{matrix} 1277.33\pm0.11 \\ 1277.3\pm0.15\end{matrix}$ \\  \\ 
BH ls220s40.0c & $\begin{matrix} \text{NO}\\ \text{IO} \end{matrix}$ & $\begin{matrix} 4.8\pm 3.7 \\ 1.9\pm 2.9\end{matrix}$ & $ \begin{matrix} 9.3\pm 5.6\\ 9.2\pm 4.1\end{matrix}$ & $ \begin{matrix} 45 \\ 44\end{matrix}$ & $\begin{matrix} 10.0\pm 3.1 \\ 7.7\pm 2.4\end{matrix}$ & $\begin{matrix} 7.3\pm 2.5 \\ 5.3\pm 1.6\end{matrix}$ & $ \begin{matrix} 7.4\pm 3.1 \\ 5.3\pm 2.6\end{matrix}$ & $ \begin{matrix} 2105.46\pm0.14 \\ 2105.41\pm0.19\end{matrix}$ \\  \\ 
BH ls220s40s7b2c & $\begin{matrix} \text{NO}\\ \text{IO} \end{matrix}$ & $\begin{matrix} 4.1\pm 3.7 \\ 0.1\pm 3.1\end{matrix}$& $ \begin{matrix} 9.7\pm 5.5\\ 9.6\pm 3.9\end{matrix}$ & $ \begin{matrix} 45 \\ 45\end{matrix}$ & $\begin{matrix} 10.0\pm 3.2 \\ 7.7\pm 2.4\end{matrix}$ & $\begin{matrix} 7.2\pm 2.5 \\ 5.2\pm 1.6\end{matrix}$ & $ \begin{matrix} 7.0\pm 3.0 \\ 5.2\pm 2.9\end{matrix}$ & $ \begin{matrix} 567.88\pm0.05 \\ 567.88\pm0.06\end{matrix}$ \\  \\ 
\hline\hline
\end{tabularx}
\caption{Summary of the results of the different timing methods in JUNO for each of the 18 SN models used.}\label{table:2}
\end{table}
\begin{table}[]
\begin{tabularx}{\textwidth}{l l Y Y Y Y Y Y Y}
\hline\hline\\
\multicolumn{9}{c}{HK - All Simulations}\\\\
Simulation & & $\begin{matrix}
\text{Exponential}\\
\text{Fit}\\
\text{[ms]}
\end{matrix}$ & $\begin{matrix}
\text{}\\
\text{Gauss Fit}\\
\text{[ms]}
\end{matrix}$ & $\begin{matrix}
\text{failed}\\
\text{Gauss Fit}\\
\text{[\%]}
\end{matrix}$ & $\begin{matrix}
\text{Energy}\\
\text{Threshold}\\
\text{($20\,$MeV) [ms]}
\end{matrix}$ & $\begin{matrix}
\text{}\\
\text{First IBD}\\
\text{[ms]}
\end{matrix}$ & $\begin{matrix}
\text{}\\
\text{First Bulk}\\
\text{[ms]}
\end{matrix}$ & $\begin{matrix}
\text{BH}\\
\text{Collapse}\\
\text{[ms]}
\end{matrix}$ \\
\hline\\
ls180s12.0 & $\begin{matrix} \text{NO}\\ \text{IO} \end{matrix}$ & $\begin{matrix} 5.1\pm 1.3 \\ 3.8\pm 0.9\end{matrix}$ & $ \begin{matrix} 9.4\pm 5.0\\ 8.2\pm 3.2\end{matrix}$ & $ \begin{matrix} 43 \\ 23\end{matrix}$ & $\begin{matrix} 6.3\pm 1.4 \\ 5.2\pm 1.0\end{matrix}$ & $\begin{matrix} 4.5\pm 1.0 \\ 3.7\pm 0.7\end{matrix}$ & $ \begin{matrix} 4.7\pm 1.1 \\ 3.7\pm 0.7\end{matrix}$ & $ \begin{matrix} -  \\ - \end{matrix}$ \\  \\ 
ls180s13.8 & $\begin{matrix} \text{NO}\\ \text{IO} \end{matrix}$ & $\begin{matrix} 6.2\pm 1.3 \\ 3.2\pm 0.9\end{matrix}$ & $ \begin{matrix} 9.2\pm 4.8\\ 8.4\pm 2.8\end{matrix}$ & $ \begin{matrix} 42 \\ 26\end{matrix}$ & $\begin{matrix} 6.1\pm 1.3 \\ 5.1\pm 1.0\end{matrix}$ & $\begin{matrix} 4.4\pm 1.0 \\ 3.6\pm 0.6\end{matrix}$ & $ \begin{matrix} 4.7\pm 1.1 \\ 3.6\pm 0.7\end{matrix}$ & $ \begin{matrix} -  \\ - \end{matrix}$ \\  \\ 
ls180s15s7b2 & $\begin{matrix} \text{NO}\\ \text{IO} \end{matrix}$ & $\begin{matrix} 4.6\pm 1.3 \\ 3.1\pm 1.0\end{matrix}$ & $ \begin{matrix} 9.5\pm 4.7\\ 8.3\pm 2.9\end{matrix}$ & $ \begin{matrix} 44 \\ 26\end{matrix}$ & $\begin{matrix} 6.4\pm 1.4 \\ 5.3\pm 0.9\end{matrix}$ & $\begin{matrix} 4.5\pm 1.0 \\ 3.7\pm 0.7\end{matrix}$ & $ \begin{matrix} 4.6\pm 1.1 \\ 3.7\pm 0.7\end{matrix}$ & $ \begin{matrix} -  \\ - \end{matrix}$ \\  \\ 
ls180s17.8 & $\begin{matrix} \text{NO}\\ \text{IO} \end{matrix}$ & $\begin{matrix} 6.1\pm 1.3 \\ 2.6\pm 0.9\end{matrix}$ & $ \begin{matrix} 9.3\pm 4.4\\ 8.5\pm 2.5\end{matrix}$ & $ \begin{matrix} 40 \\ 26\end{matrix}$ & $\begin{matrix} 6.1\pm 1.4 \\ 5.1\pm 1.0\end{matrix}$ & $\begin{matrix} 4.4\pm 1.0 \\ 3.6\pm 0.6\end{matrix}$ & $ \begin{matrix} 4.7\pm 1.1 \\ 3.6\pm 0.7\end{matrix}$ & $ \begin{matrix} -  \\ - \end{matrix}$ \\  \\ 
ls180s20.0 & $\begin{matrix} \text{NO}\\ \text{IO} \end{matrix}$ & $\begin{matrix} 5.6\pm 1.2 \\ 2.8\pm 0.9\end{matrix}$ & $ \begin{matrix} 9.4\pm 4.5\\ 8.4\pm 2.7\end{matrix}$ & $ \begin{matrix} 42 \\ 26\end{matrix}$ & $\begin{matrix} 6.2\pm 1.4 \\ 5.1\pm 0.9\end{matrix}$ & $\begin{matrix} 4.4\pm 1.0 \\ 3.6\pm 0.7\end{matrix}$ & $ \begin{matrix} 4.7\pm 1.0 \\ 3.6\pm 0.7\end{matrix}$ & $ \begin{matrix} -  \\ - \end{matrix}$ \\  \\ 
ls180s25.0 & $\begin{matrix} \text{NO}\\ \text{IO} \end{matrix}$ & $\begin{matrix} 6.4\pm 1.1 \\ 3.0\pm 0.8\end{matrix}$ & $ \begin{matrix} 9.3\pm 4.4\\ 9.1\pm 2.7\end{matrix}$ & $ \begin{matrix} 43 \\ 28\end{matrix}$ & $\begin{matrix} 5.9\pm 1.2 \\ 4.9\pm 1.0\end{matrix}$ & $\begin{matrix} 4.2\pm 0.9 \\ 3.4\pm 0.6\end{matrix}$ & $ \begin{matrix} 4.7\pm 1.0 \\ 3.4\pm 0.7\end{matrix}$ & $ \begin{matrix} -  \\ - \end{matrix}$ \\  \\ 
ls180s35.0 & $\begin{matrix} \text{NO}\\ \text{IO} \end{matrix}$ & $\begin{matrix} 6.4\pm 1.2 \\ 3.2\pm 0.9\end{matrix}$ & $ \begin{matrix} 9.4\pm 4.4\\ 8.7\pm 2.8\end{matrix}$ & $ \begin{matrix} 42 \\ 26\end{matrix}$ & $\begin{matrix} 6.0\pm 1.2 \\ 5.0\pm 0.9\end{matrix}$ & $\begin{matrix} 4.3\pm 1.0 \\ 3.5\pm 0.6\end{matrix}$ & $ \begin{matrix} 4.7\pm 1.0 \\ 3.5\pm 0.7\end{matrix}$ & $ \begin{matrix} -  \\ - \end{matrix}$ \\  \\ 
ls180s36.0 & $\begin{matrix} \text{NO}\\ \text{IO} \end{matrix}$ & $\begin{matrix} 6.1\pm 1.2 \\ 3.2\pm 0.9\end{matrix}$ & $ \begin{matrix} 9.4\pm 4.3\\ 8.8\pm 2.9\end{matrix}$ & $ \begin{matrix} 42 \\ 26\end{matrix}$ & $\begin{matrix} 6.0\pm 1.3 \\ 5.0\pm 0.9\end{matrix}$ & $\begin{matrix} 4.3\pm 0.9 \\ 3.5\pm 0.6\end{matrix}$ & $ \begin{matrix} 4.7\pm 1.0 \\ 3.5\pm 0.7\end{matrix}$ & $ \begin{matrix} -  \\ - \end{matrix}$ \\  \\ 
ls180s40.0 & $\begin{matrix} \text{NO}\\ \text{IO} \end{matrix}$ & $\begin{matrix} 6.3\pm 1.2 \\ 3.1\pm 0.8\end{matrix}$ & $ \begin{matrix} 9.5\pm 4.4\\ 9.0\pm 2.8\end{matrix}$ & $ \begin{matrix} 42 \\ 26\end{matrix}$ & $\begin{matrix} 5.9\pm 1.2 \\ 4.9\pm 0.9\end{matrix}$ & $\begin{matrix} 4.3\pm 0.9 \\ 3.5\pm 0.6\end{matrix}$ & $ \begin{matrix} 4.7\pm 1.0 \\ 3.5\pm 0.6\end{matrix}$ & $ \begin{matrix} -  \\ - \end{matrix}$ \\  \\ 
ls220s11.2 & $\begin{matrix} \text{NO}\\ \text{IO} \end{matrix}$ & $\begin{matrix} 5.2\pm 1.4 \\ 3.6\pm 0.9\end{matrix}$ & $ \begin{matrix} 9.1\pm 4.9\\ 8.1\pm 3.0\end{matrix}$ & $ \begin{matrix} 42 \\ 21\end{matrix}$ & $\begin{matrix} 6.3\pm 1.4 \\ 5.2\pm 1.0\end{matrix}$ & $\begin{matrix} 4.6\pm 1.0 \\ 3.8\pm 0.6\end{matrix}$ & $ \begin{matrix} 4.7\pm 1.1 \\ 3.8\pm 0.7\end{matrix}$ & $ \begin{matrix} -  \\ - \end{matrix}$ \\  \\ 
ls220s13.8 & $\begin{matrix} \text{NO}\\ \text{IO} \end{matrix}$ & $\begin{matrix} 6.3\pm 1.3 \\ 2.9\pm 0.9\end{matrix}$ & $ \begin{matrix} 9.1\pm 4.5\\ 8.2\pm 2.8\end{matrix}$ & $ \begin{matrix} 41 \\ 26\end{matrix}$ & $\begin{matrix} 6.0\pm 1.3 \\ 5.0\pm 1.1\end{matrix}$ & $\begin{matrix} 4.4\pm 0.9 \\ 3.6\pm 0.6\end{matrix}$ & $ \begin{matrix} 4.7\pm 1.0 \\ 3.6\pm 0.7\end{matrix}$ & $ \begin{matrix} -  \\ - \end{matrix}$ \\  \\ 
ls220s20.6 & $\begin{matrix} \text{NO}\\ \text{IO} \end{matrix}$ & $\begin{matrix} 6.2\pm 1.2 \\ 3.1\pm 0.9\end{matrix}$ & $ \begin{matrix} 9.3\pm 4.5\\ 8.9\pm 3.0\end{matrix}$ & $ \begin{matrix} 43 \\ 27\end{matrix}$ & $\begin{matrix} 5.8\pm 1.2 \\ 4.8\pm 0.9\end{matrix}$ & $\begin{matrix} 4.3\pm 0.9 \\ 3.5\pm 0.5\end{matrix}$ & $ \begin{matrix} 4.5\pm 1.0 \\ 3.5\pm 0.6\end{matrix}$ & $ \begin{matrix} -  \\ - \end{matrix}$ \\  \\ 
ls220s35.0 & $\begin{matrix} \text{NO}\\ \text{IO} \end{matrix}$ & $\begin{matrix} 6.4\pm 1.3 \\ 3.1\pm 0.9\end{matrix}$ & $ \begin{matrix} 9.3\pm 4.5\\ 8.6\pm 2.8\end{matrix}$ & $ \begin{matrix} 41 \\ 26\end{matrix}$ & $\begin{matrix} 5.9\pm 1.3 \\ 4.9\pm 1.0\end{matrix}$ & $\begin{matrix} 4.3\pm 0.9 \\ 3.6\pm 0.6\end{matrix}$ & $ \begin{matrix} 4.6\pm 1.0 \\ 3.6\pm 0.6\end{matrix}$ & $ \begin{matrix} -  \\ - \end{matrix}$ \\  \\ 
ls220s36.0 & $\begin{matrix} \text{NO}\\ \text{IO} \end{matrix}$ & $\begin{matrix} 6.2\pm 1.3 \\ 3.0\pm 0.9\end{matrix}$ & $ \begin{matrix} 9.3\pm 4.6\\ 8.6\pm 3.0\end{matrix}$ & $ \begin{matrix} 41 \\ 27\end{matrix}$ & $\begin{matrix} 5.9\pm 1.3 \\ 4.9\pm 0.9\end{matrix}$ & $\begin{matrix} 4.3\pm 0.9 \\ 3.6\pm 0.5\end{matrix}$ & $ \begin{matrix} 4.6\pm 1.0 \\ 3.6\pm 0.6\end{matrix}$ & $ \begin{matrix} -  \\ - \end{matrix}$ \\  \\ 
BH ls180s40s7b2 & $\begin{matrix} \text{NO}\\ \text{IO} \end{matrix}$ & $\begin{matrix} 4.7\pm 1.0 \\ 1.4\pm 0.8\end{matrix}$ & $ \begin{matrix} 9.7\pm 4.1\\ 9.1\pm 2.4\end{matrix}$ & $ \begin{matrix} 40 \\ 22\end{matrix}$ & $\begin{matrix} 5.4\pm 1.0 \\ 4.6\pm 0.7\end{matrix}$ & $\begin{matrix} 4.2\pm 0.8 \\ 3.5\pm 0.5\end{matrix}$ & $ \begin{matrix} 4.3\pm 0.8 \\ 3.5\pm 0.6\end{matrix}$ & $ \begin{matrix} 434.973\pm0.008 \\ 434.973\pm0.008\end{matrix}$ \\  \\ 
BH ls220s25.0c & $\begin{matrix} \text{NO}\\ \text{IO} \end{matrix}$ & $\begin{matrix} 5.8\pm 1.1 \\ 2.9\pm 0.9\end{matrix}$ & $ \begin{matrix} 9.1\pm 4.7\\ 8.5\pm 3.3\end{matrix}$ & $ \begin{matrix} 43 \\ 29\end{matrix}$ & $\begin{matrix} 5.2\pm 1.2 \\ 4.2\pm 1.0\end{matrix}$ & $\begin{matrix} 3.7\pm 0.8 \\ 3.0\pm 0.5\end{matrix}$ & $ \begin{matrix} 4.0\pm 1.0 \\ 3.0\pm 0.6\end{matrix}$ & $ \begin{matrix} 1277.432\pm0.016 \\
1277.426\pm0.022\end{matrix}$ \\  \\ 
BH ls220s40.0c & $\begin{matrix} \text{NO}\\ \text{IO} \end{matrix}$ & $\begin{matrix} 6.0\pm 1.2 \\ 3.3\pm 0.9\end{matrix}$ & $ \begin{matrix} 9.3\pm 4.4\\ 8.8\pm 3.0\end{matrix}$ & $ \begin{matrix} 44 \\ 27\end{matrix}$ & $\begin{matrix} 5.8\pm 1.2 \\ 4.8\pm 0.9\end{matrix}$ & $\begin{matrix} 4.2\pm 0.8 \\ 3.6\pm 0.5\end{matrix}$ & $ \begin{matrix} 4.5\pm 0.9 \\ 3.6\pm 0.6\end{matrix}$ & $ \begin{matrix} 2105.583\pm0.018 \\ 2105.573\pm0.029\end{matrix}$ \\  \\ 
BH ls220s40s7b2c & $\begin{matrix} \text{NO}\\ \text{IO} \end{matrix}$ & $\begin{matrix} 5.4\pm 1.1 \\ 1.6\pm 0.9\end{matrix}$ & $ \begin{matrix} 9.8\pm 4.1\\ 9.3\pm 2.5\end{matrix}$ & $ \begin{matrix} 42 \\ 24\end{matrix}$ & $\begin{matrix} 5.7\pm 1.3 \\ 4.7\pm 0.8\end{matrix}$ & $\begin{matrix} 4.1\pm 0.8 \\ 3.5\pm 0.5\end{matrix}$ & $ \begin{matrix} 4.3\pm 0.9 \\ 3.5\pm 0.7\end{matrix}$ & $ \begin{matrix} 567.924\pm0.008 \\ 567.924\pm0.008\end{matrix}$ \\  \\

\hline\hline
\end{tabularx}
\caption{Summary of the results of the different timing methods in Hyper-Kamiokande for each of the 18 SN models used.}\label{table:3}
\end{table}
\begin{table}[]
\begin{tabularx}{0.45 \textwidth}{l l Y Y}
\hline\hline\\
 & & IC & IC Gen2\\\\
Simulation & & $\begin{matrix}
\text{Exponential Fit}\\
\text{[ms]}
\end{matrix}$ & $\begin{matrix}
\text{Exponential Fit}\\
\text{[ms]}
\end{matrix}$\\
\hline\\
ls180s12.0 & $\begin{matrix} \text{NO}\\ \text{IO} \end{matrix}$ & $\begin{matrix} 4.7\pm 1.4 \\ 4.2\pm 1.1\end{matrix}$ & $\begin{matrix} 4.6\pm 0.9 \\ 4.3\pm 0.8\end{matrix}$ \\  \\ 
ls180s13.8 & $\begin{matrix} \text{NO}\\ \text{IO} \end{matrix}$ & $\begin{matrix} 5.7\pm 1.1 \\ 3.6\pm 1.0\end{matrix}$ & $\begin{matrix} 5.6\pm 0.7 \\ 3.5\pm 0.6\end{matrix}$ \\  \\ 
ls180s15s7b2 & $\begin{matrix} \text{NO}\\ \text{IO} \end{matrix}$ & $\begin{matrix} 4.2\pm 1.3 \\ 3.2\pm 1.1\end{matrix}$ & $\begin{matrix} 4.1\pm 0.9 \\ 3.2\pm 0.7\end{matrix}$ \\  \\ 
ls180s17.8 & $\begin{matrix} \text{NO}\\ \text{IO} \end{matrix}$ & $\begin{matrix} 5.6\pm 1.1 \\ 2.9\pm 0.9\end{matrix}$ & $\begin{matrix} 5.5\pm 0.7 \\ 2.9\pm 0.6\end{matrix}$ \\  \\ 
ls180s20.0 & $\begin{matrix} \text{NO}\\ \text{IO} \end{matrix}$ & $\begin{matrix} 5.1\pm 1.1 \\ 3.0\pm 1.0\end{matrix}$ & $\begin{matrix} 5.1\pm 0.7 \\ 2.9\pm 0.6\end{matrix}$ \\  \\ 
ls180s25.0 & $\begin{matrix} \text{NO}\\ \text{IO} \end{matrix}$ & $\begin{matrix} 5.8\pm 0.9 \\ 3.1\pm 0.8\end{matrix}$ & $\begin{matrix} 5.8\pm 0.6 \\ 3.0\pm 0.5\end{matrix}$ \\  \\ 
ls180s35.0 & $\begin{matrix} \text{NO}\\ \text{IO} \end{matrix}$ & $\begin{matrix} 5.8\pm 1.0 \\ 3.4\pm 0.9\end{matrix}$ & $\begin{matrix} 5.8\pm 0.7 \\ 3.4\pm 0.6\end{matrix}$ \\  \\ 
ls180s36.0 & $\begin{matrix} \text{NO}\\ \text{IO} \end{matrix}$ & $\begin{matrix} 5.5\pm 1.0 \\ 3.3\pm 0.9\end{matrix}$ & $\begin{matrix} 5.5\pm 0.7 \\ 3.3\pm 0.6\end{matrix}$ \\  \\ 
ls180s40.0 & $\begin{matrix} \text{NO}\\ \text{IO} \end{matrix}$ & $\begin{matrix} 5.7\pm 1.0 \\ 3.3\pm 0.8\end{matrix}$ & $\begin{matrix} 5.7\pm 0.6 \\ 3.2\pm 0.5\end{matrix}$ \\  \\ 
ls220s11.2 & $\begin{matrix} \text{NO}\\ \text{IO} \end{matrix}$ & $\begin{matrix} 4.9\pm 1.5 \\ 3.9\pm 1.2\end{matrix}$ & $\begin{matrix} 4.8\pm 0.9 \\ 4.2\pm 0.8\end{matrix}$ \\  \\ 
ls220s13.8 & $\begin{matrix} \text{NO}\\ \text{IO} \end{matrix}$ & $\begin{matrix} 5.9\pm 1.2 \\ 3.4\pm 1.0\end{matrix}$ & $\begin{matrix} 5.8\pm 0.7 \\ 3.3\pm 0.6\end{matrix}$ \\  \\ 
ls220s20.6 & $\begin{matrix} \text{NO}\\ \text{IO} \end{matrix}$ & $\begin{matrix} 5.7\pm 1.0 \\ 3.3\pm 0.9\end{matrix}$ & $\begin{matrix} 5.7\pm 0.6 \\ 3.2\pm 0.6\end{matrix}$ \\  \\ 
ls220s35.0 & $\begin{matrix} \text{NO}\\ \text{IO} \end{matrix}$ & $\begin{matrix} 5.9\pm 1.1 \\ 3.4\pm 0.9\end{matrix}$ & $\begin{matrix} 5.9\pm 0.7 \\ 3.3\pm 0.6\end{matrix}$ \\  \\ 
ls220s36.0 & $\begin{matrix} \text{NO}\\ \text{IO} \end{matrix}$ & $\begin{matrix} 5.6\pm 1.1 \\ 3.2\pm 0.9\end{matrix}$ & $\begin{matrix} 5.6\pm 0.7 \\ 3.1\pm 0.6\end{matrix}$ \\  \\ 
BH ls180s40s7b2 & $\begin{matrix} \text{NO}\\ \text{IO} \end{matrix}$ & $\begin{matrix} 4.0\pm 0.8 \\ 1.2\pm 0.6\end{matrix}$ & $\begin{matrix} 3.9\pm 0.5 \\ 1.1\pm 0.4\end{matrix}$ \\  \\ 
BH ls220s25.0c & $\begin{matrix} \text{NO}\\ \text{IO} \end{matrix}$ & $\begin{matrix} 5.2\pm 0.9 \\ 3.0\pm 0.8\end{matrix}$ & $\begin{matrix} 5.2\pm 0.6 \\ 2.9\pm 0.5\end{matrix}$ \\  \\ 
BH ls220s40.0c & $\begin{matrix} \text{NO}\\ \text{IO} \end{matrix}$ & $\begin{matrix} 5.5\pm 0.9 \\ 3.5\pm 0.8\end{matrix}$ & $\begin{matrix} 5.5\pm 0.6 \\ 3.3\pm 0.5\end{matrix}$ \\  \\ 
BH ls220s40s7b2c & $\begin{matrix} \text{NO}\\ \text{IO} \end{matrix}$ & $\begin{matrix} 5.0\pm 0.9 \\ 1.6\pm 0.8\end{matrix}$ & $\begin{matrix} 4.9\pm 0.6 \\ 1.5\pm 0.5\end{matrix}$ \\  \\ 
\hline\hline
\end{tabularx}
\caption{Summary of the results of the different timing methods in IceCube and its potential IceCube Gen 2 upgrade for each of the 18 SN models used. The time is measured relative to core bounce.}\label{table:4}
\end{table}
\newpage
\clearpage
\section{Simulations with Massive Neutrinos}\label{App. B}
Here we show some selected simulations with neutrinos of different masses in TABLE~\ref{tab:massive_simulations}. The simulations were done for the ls180s12.0 and BH ls220s40s7b2c models.
\begin{table}[hb]
\begin{tabularx}{\textwidth}{l l Y Y Y Y Y Y Y}
\hline\hline\\
\multicolumn{9}{c}{HK - Different Neutrino Masses}\\\\
Simulation & & $\begin{matrix}
\text{Exponential}\\
\text{Fit}\\
\text{[ms]}
\end{matrix}$ & $\begin{matrix}
\text{}\\
\text{Gauss Fit}\\
\text{[ms]}
\end{matrix}$ & $\begin{matrix}
\text{failed}\\
\text{Gauss Fit}\\
\text{[\%]}
\end{matrix}$ & $\begin{matrix}
\text{Energy}\\
\text{Threshold}\\
\text{($20\,$MeV) [ms]}
\end{matrix}$ & $\begin{matrix}
\text{}\\
\text{First IBD}\\
\text{[ms]}
\end{matrix}$ & $\begin{matrix}
\text{}\\
\text{First Bulk}\\
\text{[ms]}
\end{matrix}$ & $\begin{matrix}
\text{BH}\\
\text{Collapse}\\
\text{[ms]}
\end{matrix}$ \\
\hline
\\
ls180s12.0 & $\begin{matrix} \text{NO}\\ \text{IO} \end{matrix}$ & $\begin{matrix} 5.1\pm 1.3 \\ 3.8\pm 0.9\end{matrix}$ & $ \begin{matrix} 9.4\pm 5.0\\ 8.2\pm 3.2\end{matrix}$ & $ \begin{matrix} 43 \\ 23\end{matrix}$ & $\begin{matrix} 6.3\pm 1.4 \\ 5.2\pm 1.0\end{matrix}$ & $\begin{matrix} 4.5\pm 1.0 \\ 3.7\pm 0.7\end{matrix}$ & $ \begin{matrix} 4.7\pm 1.1 \\ 3.7\pm 0.7\end{matrix}$ & $ \begin{matrix} -  \\ - \end{matrix}$ \\[0.4cm]

BH ls220s40s7b2c & $\begin{matrix} \text{NO}\\ \text{IO} \end{matrix}$ & $\begin{matrix} 5.4\pm 1.1 \\ 1.6\pm 0.9\end{matrix}$ & $ \begin{matrix} 9.8\pm 4.1\\ 9.3\pm 2.5\end{matrix}$ & $ \begin{matrix} 42 \\ 24\end{matrix}$ & $\begin{matrix} 5.7\pm 1.3 \\ 4.7\pm 0.8\end{matrix}$ & $\begin{matrix} 4.1\pm 0.8 \\ 3.5\pm 0.5\end{matrix}$ & $ \begin{matrix} 4.3\pm 0.9 \\ 3.5\pm 0.7\end{matrix}$ & $ \begin{matrix} 567.924\pm0.008 \\ 567.924\pm0.008\end{matrix}$ \\  \\  
 & \multicolumn{7}{c}{$0.05\,$eV Neutrinos}\\\\
ls180s12.0 & $\begin{matrix} \text{NO}\\ \text{IO} \end{matrix}$ & $\begin{matrix} 5.0\pm 1.3 \\ 3.8\pm 0.9\end{matrix}$ & $ \begin{matrix} 9.0\pm 4.9\\ 8.2\pm 2.9\end{matrix}$ & $ \begin{matrix} 43 \\ 23\end{matrix}$ & $\begin{matrix} 6.3\pm 1.4 \\ 5.3\pm 1.0\end{matrix}$ & $\begin{matrix} 4.6\pm 1.0 \\ 3.7\pm 0.7\end{matrix}$ & $ \begin{matrix} 4.7\pm 1.0 \\ 3.7\pm 0.7\end{matrix}$ & $ \begin{matrix} -  \\ - \end{matrix}$ \\[0.4cm]  

BH ls220s40s7b2c & $\begin{matrix} \text{NO}\\ \text{IO} \end{matrix}$ & $\begin{matrix} 5.3\pm 1.1 \\ 1.5\pm 0.9\end{matrix}$ & $ \begin{matrix} 9.8\pm 4.5\\ 9.3\pm 2.3\end{matrix}$ & $ \begin{matrix} 42 \\ 23\end{matrix}$ & $\begin{matrix} 5.7\pm 1.2 \\ 4.8\pm 0.9\end{matrix}$ & $\begin{matrix} 4.2\pm 0.8 \\ 3.5\pm 0.5\end{matrix}$ & $ \begin{matrix} 4.3\pm 0.9 \\ 3.5\pm 0.7\end{matrix}$ & $ \begin{matrix} 567.926\pm0.008 \\ 567.925\pm0.008\end{matrix}$ \\  \\ 

 & \multicolumn{7}{c}{$0.1\,$eV Neutrinos}\\\\
ls180s12.0 & $\begin{matrix} \text{NO}\\ \text{IO} \end{matrix}$ & $\begin{matrix} 5.1\pm 1.3 \\ 3.7\pm 0.9\end{matrix}$ & $ \begin{matrix} 9.5\pm 4.8\\ 8.2\pm 2.7\end{matrix}$ & $ \begin{matrix} 45 \\ 24\end{matrix}$ & $\begin{matrix} 6.3\pm 1.4 \\ 5.3\pm 1.0\end{matrix}$ & $\begin{matrix} 4.5\pm 1.1 \\ 3.7\pm 0.7\end{matrix}$ & $ \begin{matrix} 4.7\pm 1.1 \\ 3.7\pm 0.7\end{matrix}$ & $ \begin{matrix} -  \\ - \end{matrix}$ \\[0.4cm]

BH ls220s40s7b2c & $\begin{matrix} \text{NO}\\ \text{IO} \end{matrix}$ & $\begin{matrix} 5.5\pm 1.2 \\ 1.6\pm 0.9\end{matrix}$ & $ \begin{matrix} 9.6\pm 4.1\\ 9.3\pm 2.7\end{matrix}$ & $ \begin{matrix} 39 \\ 21\end{matrix}$ & $\begin{matrix} 5.7\pm 1.2 \\ 4.7\pm 0.8\end{matrix}$ & $\begin{matrix} 4.1\pm 0.8 \\ 3.5\pm 0.5\end{matrix}$ & $ \begin{matrix} 4.3\pm 0.9 \\ 3.5\pm 0.7\end{matrix}$ & $ \begin{matrix} 567.931\pm0.011 \\ 567.93\pm0.011\end{matrix}$ \\  \\ 

 & \multicolumn{7}{c}{$0.5\,$eV Neutrinos}\\\\
ls180s12.0 & $\begin{matrix} \text{NO}\\ \text{IO} \end{matrix}$ & $\begin{matrix} 5.4\pm 1.3 \\ 4.1\pm 0.8\end{matrix}$ & $ \begin{matrix} 9.2\pm 4.9\\ 8.6\pm 2.7\end{matrix}$ & $ \begin{matrix} 45 \\ 23\end{matrix}$ & $\begin{matrix} 6.5\pm 1.4 \\ 5.4\pm 0.9\end{matrix}$ & $\begin{matrix} 5.1\pm 1.0 \\ 4.3\pm 0.6\end{matrix}$ & $ \begin{matrix} 5.1\pm 1.0 \\ 4.3\pm 0.7\end{matrix}$ & $ \begin{matrix} -  \\ - \end{matrix}$ \\[0.4cm]

BH ls220s40s7b2c & $\begin{matrix} \text{NO}\\ \text{IO} \end{matrix}$ & $\begin{matrix} 5.7\pm 1.1 \\ 2.0\pm 0.8\end{matrix}$ & $ \begin{matrix} 9.9\pm 3.7\\ 9.7\pm 2.4\end{matrix}$ & $ \begin{matrix} 41 \\ 25\end{matrix}$ & $\begin{matrix} 5.9\pm 1.2 \\ 4.9\pm 0.7\end{matrix}$ & $\begin{matrix} 4.6\pm 0.8 \\ 4.1\pm 0.5\end{matrix}$ & $ \begin{matrix} 4.8\pm 0.9 \\ 4.1\pm 0.7\end{matrix}$ & $ \begin{matrix} 568.63\pm0.32 \\ 568.51\pm0.32\end{matrix}$ \\  \\ 

 & \multicolumn{7}{c}{$1\,$eV Neutrinos}\\\\
ls180s12.0 & $\begin{matrix} \text{NO}\\ \text{IO} \end{matrix}$ & $\begin{matrix} 6.4\pm 1.3 \\ 5.3\pm 0.8\end{matrix}$ & $ \begin{matrix} 9.7\pm 4.8\\ 9.0\pm 4.1\end{matrix}$ & $ \begin{matrix} 44 \\ 23\end{matrix}$ & $\begin{matrix} 7.1\pm 1.3 \\ 6.0\pm 0.9\end{matrix}$ & $\begin{matrix} 6.4\pm 0.9 \\ 5.6\pm 0.6\end{matrix}$ & $ \begin{matrix} 6.3\pm 1.1 \\ 5.6\pm 0.6\end{matrix}$ & $ \begin{matrix} -  \\ - \end{matrix}$ \\[0.4cm]

BH ls220s40s7b2c & $\begin{matrix} \text{NO}\\ \text{IO} \end{matrix}$ & $\begin{matrix} 6.8\pm 1.1 \\ 3.1\pm 0.9\end{matrix}$ & $ \begin{matrix} 10.3\pm 3.9\\ 10.9\pm 2.3\end{matrix}$ & $ \begin{matrix} 44 \\ 30\end{matrix}$ & $\begin{matrix} 6.5\pm 1.2 \\ 5.6\pm 0.7\end{matrix}$ & $\begin{matrix} 5.9\pm 0.8 \\ 5.2\pm 0.5\end{matrix}$ & $ \begin{matrix} 6.0\pm 0.9 \\ 5.2\pm 0.7\end{matrix}$ & $ \begin{matrix} 572.1\pm1.1 \\ 571.8\pm1.3\end{matrix}$ \\  \\ 

 & \multicolumn{7}{c}{$2\,$eV Neutrinos}\\\\
ls180s12.0 & $\begin{matrix} \text{NO}\\ \text{IO} \end{matrix}$ & $\begin{matrix} 10.1\pm 1.3 \\ 9.0\pm 1.0\end{matrix}$ & $ \begin{matrix} 7.8\pm 6.5\\ 10.5\pm 6.5\end{matrix}$ & $ \begin{matrix} 36 \\ 34\end{matrix}$ & $\begin{matrix} 9.3\pm 1.3 \\ 8.2\pm 0.9\end{matrix}$ & $\begin{matrix} 9.3\pm 1.2 \\ 8.2\pm 0.9\end{matrix}$ & $ \begin{matrix} 9.7\pm 1.2 \\ 8.2\pm 0.9\end{matrix}$ & $ \begin{matrix} -  \\ - \end{matrix}$ \\[0.4cm] 

BH ls220s40s7b2c & $\begin{matrix} \text{NO}\\ \text{IO} \end{matrix}$ & $\begin{matrix} 10.7\pm 1.2 \\ 6.9\pm 0.9\end{matrix}$ & $ \begin{matrix} 8.3\pm 7.0\\ 12.4\pm 5.4\end{matrix}$ & $ \begin{matrix} 42 \\ 38\end{matrix}$ & $\begin{matrix} 8.7\pm 1.2 \\ 7.7\pm 0.8\end{matrix}$ & $\begin{matrix} 8.7\pm 1.1 \\ 7.7\pm 0.8\end{matrix}$ & $ \begin{matrix} 9.5\pm 1.2 \\ 7.7\pm 0.8\end{matrix}$ & $ \begin{matrix} 589.3\pm3.5 \\ 588.6\pm4.3\end{matrix}$ \\  \\ 

\hline\hline
\end{tabularx}
\caption{Summary of the results of the different timing methods in Hyper-Kamiokande assuming different neutrino masses and their influence on the time of flight. The time is measured relative to core bounce. The results are shown for two different SN models as examples. For each SN model and neutrino mass 1000 Simulations have been performed.}\label{tab:massive_simulations}
\end{table}

\FloatBarrier
\newpage
\twocolumngrid
\section*{}
\qquad
\bibliography{revtexbib}
\end{document}